\documentclass[11pt,a4paper]{article}
\pdfoutput=1

\usepackage[width=6.1in, height=8.9in]{geometry}
\usepackage{amsmath, amssymb, graphicx, multirow}

\title{\bf The four fixed points of scale invariant single field cosmological models}

\author{BingKan Xue\\
Department of Physics, Princeton University,\\
Princeton, New Jersey 08544, USA\\
bxue@princeton.edu}

\date{\today}

\begin{document}

\maketitle

\begin{abstract}
We introduce a new set of flow parameters to describe the time dependence of the equation of state and the speed of sound in single field cosmological models. A scale invariant power spectrum is produced if these flow parameters satisfy specific dynamical equations. We analyze the flow of these parameters and find four types of fixed points that encompass all known single field models. Moreover, near each fixed point we uncover new models where the scale invariance of the power spectrum relies on having simultaneously time varying speed of sound and equation of state. We describe several distinctive new models and discuss constraints from strong coupling and superluminality.
\end{abstract}

\tableofcontents

\newpage
\section{Introduction}

Our Universe is observed to be homogeneous and isotropic at large distances from super-galactic scales ($\sim 1$ Mpc) to the whole visible Universe ($\sim 10^4$ Mpc). Moreover, the power spectrum of the primordial density fluctuations on such large scales is measured to be nearly scale invariant. According to the current view, these perturbations arose as quantum fluctuations in the early Universe, then left the horizon before the standard expansion phase, only to recently reenter. The evolution of the adiabatic fluctuations in the early Universe are determined by the equation of state $w$ and the speed of sound $c_s$. The question is, which types of early cosmic evolution $w(t), c_s(t)$ would give rise to the observed 10 e-folds of nearly scale invariant perturbations? 

In a common class of models for generating scale invariant perturbations, the Universe is dominated by a scalar field with the canonical kinetic term and certain potential term. In this case $c_s$ equals $1$, and the evolution of the scale factor $a$, hence $H \equiv \dot{a}/a$ and $\epsilon \equiv -\dot{H}/H^2 = \frac{3}{2}(1+w)$, is determined by the form of the potential. Two well known such models are the inflation and the adiabatic ekpyrosis, which can both produce a scale invariant power spectrum. In the inflationary scenario \cite{Guth:1980zm, Linde:1981mu, Albrecht:1982wi}, $\epsilon$ stays nearly constant and close to $0$, while the scale factor $a$ grows exponentially, pushing a wide range of length scales outside the horizon. In the adiabatic ekpyrotic scenario \cite{Khoury:2009my, Khoury:2011ii}, however, the scale factor $a$ contracts slowly while $\epsilon$ increases rapidly, so that the horizon scale shrinks and leaves a wide range of perturbation modes outside. The scale invariant modes created in a contraction phase can be carried on to the standard expansion phase through a cosmic bounce \cite{Tolley:2003nx, Turok:2004gb, McFadden:2005mq}.

For a constant $c_s$, the scale invariance of the power spectrum fully constrains the time dependence of the scale factor $a(t)$ \cite{Khoury:2010gw, Baumann:2011dt}. In addition to inflation and adiabatic ekpyrosis, there is a third type of solution in which the universe goes through an ``apex'' \cite{Khoury:2010gw}, a slow transition from expansion to contraction. Similar to adiabatic ekpyrosis, this scenario also depends on a rapidly changing $\epsilon$ while the scale factor $a$ is nearly constant. One difference is that here the scale invariant modes are generated during the slow expansion phase before the apex, instead of during a slow contraction phase.

More generally, the speed of sound $c_s$ may vary with time, as in single field models with a noncanonical kinetic term. Several models involving a time varying $c_s$ have been considered. In the tachyacoustic expansion scenario \cite{ArmendarizPicon:2006if, Piao:2006ja, Magueijo:2008pm, Bessada:2009ns}, $\epsilon$ is assumed to be constant, and the scale invariant power spectrum is produced by having a suitable time dependence of the speed of sound $c_s$. Another example is considered in \cite{Joyce:2011kh} where $c_s$ is taken to be proportional to $\epsilon$, so that the three-point function of the curvature perturbation is also scale invariant.

With a time dependent $c_s$, the requirement of a scale invariant power spectrum alone does not fully determine the cosmic evolution. There exist infinitely many combinations of $a(t)$ and $c_s(t)$ that can all produce scale invariant perturbations. We would like to know if these different scenarios can be described and categorized into distinctive classes of cosmological models.

In this paper we present a general scheme to describe single field cosmological models by a new set of flow parameters. Any single field model can be represented by a point or a trajectory in this parameter space. The requirement of a scale invariant power spectrum constrains the dynamics of the flow parameters and determines particular classes of trajectories. We show that all existing models correspond to nearly constant values of the flow parameters. By looking for fixed points in the parameter space, we find four general types of models that encompass inflation, adiabatic ekpyrosis, apex, and tachyacoustic expansion respectively. Furthermore, we analyze the flow lines near the fixed points in different cross sections of the parameter space. By looking for various behavior of each fixed point, we obtain new cosmological models that can produce scale invariant perturbations equally well.

Our approach is very different from the conventional way of starting from a particular scalar field Lagrangian and studing its cosmological solutions. In that approach, generally speaking, only solutions in a limited region of field phase space would yield scale invariant perturbations, corresponding to a particular type of mechanism represented by some part of the flow parameter space. In our approach, however, by tracing the flow lines in the whole parameter space, we obtain all possible cosmic evolutions that can create a scale invariant power spectrum. Each and every flow line leads to scale invariant perturbations, but different trajectories may correspond to very different scalar field models. This approach effectively avoids the limitations of particular scalar field models.

In section~\ref{sec:flow} we define the flow parameters and specify their dynamics in order to produce scale invariant perturbations. We use these parameters to study the $c_s = 1$ case and compare our results with known models. In section~\ref{sec:stat} we generalize to a time dependent speed of sound and analyze all four fixed points of the flow parameters. In particular, we uncover new scenarios and describe the corresponding new types of cosmological models. A few examples are shown in section~\ref{sec:exam}. Section~\ref{sec:cons} provides further discussions on physical constraints from strong coupling and superluminality etc. We conclude with a summary in section~\ref{sec:summ}.

\section{Flow parameters} \label{sec:flow}

Consider a single scalar degree of freedom evolving on a Friedmann-Robertson-Walker background. The adiabatic fluctuations can be studied by using the effective action for the curvature perturbation $\zeta$ \cite{Khoury:2008wj, Baumann:2011dt}. At quadratic order, the action is given by
\begin{equation} \label{eq:S2}
S_2 = M_\text{Pl}^2 \int d^3 x dy \, q^2 \big[ (\zeta')^2 - (\partial_i \zeta)^2 \big] ,
\end{equation}
where $q^2 \equiv a^2 \epsilon / c_s$ describes an effective background geometry, and $^\prime$ denotes the derivative with respect to the ``sound horizon time'' $y$, related to the physical time $t$ by $dy = (c_s / a) dt$. Normalizing the field $\zeta$ by $v \equiv \sqrt{2} q \zeta$, the quadratic action becomes
\begin{equation}
S_2 = M_\text{Pl}^2 \int d^3 x dy \, \frac{1}{2} \Big[ (v')^2 - (\partial_i v)^2 + \frac{q''}{q} v^2 \Big] .
\end{equation}

The variable $v$ can be canonically quantized in the standard way \cite{Mukhanov:1990me}. The Fourier modes $v_k$ should satisfy the equation of motion
\begin{equation} \label{eq:vk}
{v_k}'' + \Big( k^2 - \frac{q''}{q} \Big) v_k = 0 .
\end{equation}
In comparison to the comoving scale $1/k$, the inverse square root of $|q''/q|$ represents the \emph{freeze-out horizon} size. In general, if $q \propto y^n$, then $q''/q = n(n-1) / y^2$, hence the horizon size is roughly $\sim |y|$. Modes deep inside the horizon, with $k|y| \gg 1$, are effectively in a Minkowski background. Specifically, eq.~(\ref{eq:vk}) has an exact solution in terms of the Hankel function,
\begin{equation} \label{eq:hankel}
v_k (y) = \sqrt{\frac{\pi y}{4}} \, H^{(1)}_\nu (-ky) ,
\end{equation}
where $\nu = | n - \frac{1}{2} |$. This solution has been chosen to match the Minkowski vacuum state in the asymptotic past when the mode is deep inside the horizon, $v_k \sim e^{-iky} / \sqrt{2k}$ as $ky \to -\infty$. The horizon crossing happens at a later time when $k|y| \sim 1$. Accordingly, the time variable $y$ runs from $-\infty$ to $0$.

At late times when $k|y| \ll 1$, the curvature perturbation $\zeta_k = v_k / \sqrt{2} q$ becomes
\begin{equation} \label{eq:zeta_k}
\zeta_k = C_1 k^{-\nu} (-y)^{\frac{1}{2} - n - \nu} + C_2 k^{\nu} (-y)^{\frac{1}{2} - n + \nu} + \mathcal{O}(k^{2 + \nu}) ,
\end{equation}
where $C_1, C_2$ are dimensionless Taylor coefficients. Note that for $n < \frac{1}{2}$, the leading term in eq.~(\ref{eq:zeta_k}) is time independent while the second and higher terms are decaying at late times, hence $\zeta_k$ approaches constant values on large scales outside the horizon. But for $n > \frac{1}{2}$, however, the leading term keeps growing even after crossing the horizon, indicating an instability in the curvature perturbation against the homogeneous background.

The power spectrum of the curvature perturbation $\zeta$ is given by
\begin{equation} \label{eq:power}
P_\zeta = \frac{k^3}{2\pi^2} \, |\zeta_k|^2 ,
\end{equation}
where $\zeta_k$ should be evaluated at horizon crossing or when the super-horizon modes stop growing \cite{Khoury:2008wj}. In either cases, the spectral tilt is given by the leading term in (\ref{eq:zeta_k}),
\begin{equation}
n_s - 1 = 3 - 2 \nu = 3 - |2 n - 1| .
\end{equation}
For the power spectrum to be scale invariant, one needs $n = -1$ or $2$. The latter case with $n > \frac{1}{2}$ is subject to unstable growths of curvature perturbation on large scales; hence we only consider the former case with $n = -1$ in this paper.

Therefore, in order to have a scale invariant power spectrum, it suffices to find functions $a(y)$ and $c_s(y)$ such that $q = a \sqrt{\epsilon / c_s} \propto 1/(-y)$. In general, the relation $q \propto (-y)^n$ can be written as
\begin{equation} \label{eq:n}
n = \frac{d \log q}{d \log (-y)} ,
\end{equation}
where $n = -1$ in the scale invariant case. Define the following flow parameters in a similar way,
\begin{align}
p &\equiv \frac{d \log a}{d \log (-y)} = \frac{a H y}{c_s} , \label{eq:p} \\[4pt]
\tilde{s} &\equiv \frac{d \log c_s}{d \log (-y)} , \label{eq:stilde} \\[4pt]
r &\equiv \frac{d \log (a H)}{d \log (-y)} = \frac{d \log |p|}{d \log (-y)} + \tilde{s} - 1 . \label{eq:r}
\end{align}
Note that the more commonly defined parameter $s \equiv \dot{c_s} / H c_s$ can be expressed as $s = \tilde{s} / p$.

The parameter $p$ measures the expansion or contraction rate of the universe. Since $y$ goes from $-\infty$ to $0$, $p < 0$ corresponds to an expanding universe, whereas $p > 0$ corresponds to a contracting universe. The parameter $\tilde{s}$ measures the time dependence of the speed of sound $c_s$. $c_s = 1$ corresponds to $\tilde{s} = 0$; for $\tilde{s} < 0$ the speed of sound $c_s$ increases with time, whereas for $\tilde{s} > 0$ it decreases with time. The parameter $r$ measures how the comoving Hubble horizon $1/aH$ scale with the freeze-out horizon $\sqrt{|q/q''|} \sim (-y)$. When $\epsilon$ or $c_s$ varies with time, these two horizon scales do not coincide. Note that the parameter $p$ also measures the ratio between the freeze-out horizon $(-y)$ and the ``comoving sound horizon'' $c_s / a H$.

In terms of these flow parameters, the parameter $\epsilon = \frac{3}{2} (1+w)$ is given by
\begin{equation} \label{eq:epsilon}
\epsilon \equiv \frac{-\dot{H}}{H^2} = \frac{p - r}{p} .
\end{equation}
The null energy condition implies $\epsilon \geq 0$, which will be respected in this paper. To compute $n$, define the parameter
\begin{equation} \label{eq:eta}
\tilde{\eta} \equiv \frac{d \log \epsilon}{d \log (-y)} = \frac{1}{p - r} \bigg[ r (r - \tilde{s} + 1) - \frac{d r}{d \log (-y)} \bigg] .
\end{equation}
It is related to the usual slow-roll parameter $\eta \equiv \dot{\epsilon} / H \epsilon$ by $\eta = \tilde{\eta} / p$. With these parameters, it is straightforward to write $n$ in eq.~(\ref{eq:n}) as
\begin{equation} \label{eq:n=}
n = p + \frac{1}{2} (\tilde{\eta} - \tilde{s}) .
\end{equation}
To have a constant $n$, equations (\ref{eq:r}) and (\ref{eq:eta}) imply the following equations,
\begin{align}
- \frac{d p}{d \log(-y)} &= - p (r - \tilde{s} + 1) , \label{eq:flowp} \\[4pt]
- \frac{d r}{d \log(-y)} &= - p (2 p - \tilde{s} - 2 n) + 2 p r - r (r + 2 n + 1) . \label{eq:flowr}
\end{align}
On the left hand side, the derivative is with respect to the ``log time'' $-\log(-y)$, which runs from $-\infty$ to $\infty$. Given the value of $\tilde{s}$, these two equations determine the flow in the parameter space $(p, r, \tilde{s})$.

\begin{figure}[htb]
\centering
\includegraphics[width=5in]{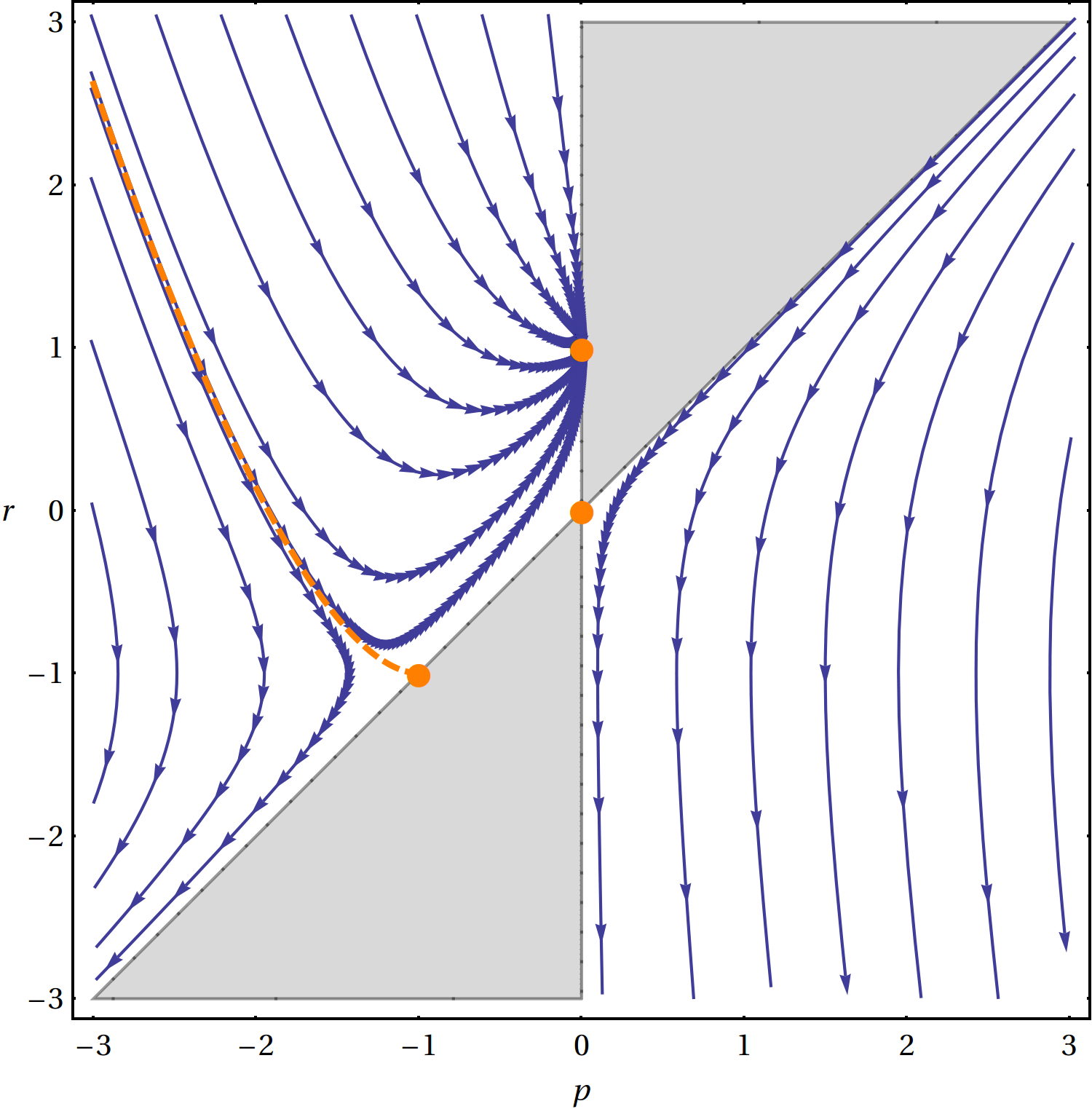}
\caption{Flow lines for $n = -1$ in the $p \, r$-plane with $\tilde{s} = 0$. The three solid circles are the fixed points representing inflation $(-1,-1)$, adiabatic ekpyrosis $(0,0)$, and apex $(0,1)$ respectively; the dashed line is a separatrix. Each line segment between adjacent arrowheads on the flow lines represents $0.1$ log time interval, or $0.1$ e-fold of scale invariant modes.} \label{fig:s=0}
\end{figure}

For example, in the case where $c_s = $ const, the parameter $\tilde{s} = 0$. In the $p \, r$-plane with $\tilde{s} = 0$, the flow lines according to eqs.~(\ref{eq:flowp}, \ref{eq:flowr}) for $n = -1$ are shown in figure~\ref{fig:s=0}.
The shaded region corresponds to $\epsilon = (p-r)/p < 0$, which is forbidden by the null energy condition. There are three fixed points at $(p,r) = (-1,-1)$, $(0,0)$, and $(0,1)$ respectively. The one at $(0,1)$ behaves as a sink where nearby trajectories converge. The one at $(0,0)$ is a saddle point where trajectories approach from both sides but eventually move away. The other saddle point at $(-1,-1)$ is attached to a separatrix which, together with the shaded region, divides the plane into three sectors. The flow lines in the bulk of each sector are qualitatively similar.

Every flow line in the figure corresponds to a cosmological model that yields scale invariant perturbations. But typically the time dependence of the flow parameters along a trajectory is very complicated. The more interesting trajectories are those that come near the fixed points. There the flow velocities given by eqs.~(\ref{eq:flowp}, \ref{eq:flowr}) approach zero, hence the flow parameters stay near those points for a sustained period in log time $-\log(-y)$. Since $-\log(-y) \sim \log k$ at horizon crossing, during that period a large number of modes exit the horizon with a scale invariant spectrum. The corresponding cosmological models can be characterized by nearly constant values of the flow parameters, which allows simple analytic descriptions of their time evolution. Indeed, the three fixed points in figure~\ref{fig:s=0} precisely correspond to the three well known types of scale invariant models with $c_s = 1$. Let us identify each of them.

The fixed point at $(-1,-1)$ corresponds to \emph{inflation} \cite{Guth:1980zm, Linde:1981mu, Albrecht:1982wi}. On the point $(-1,-1)$ the parameter $\epsilon = 0$, which corresponds to the false vacuum inflation where $H =$ const until the universe tunnels to the true vacuum. In that case $a H = 1 / (-\tau)$, where $\tau$ is the conformal time that equals the sound horizon time $y$ for $c_s = 1$, hence $p = a H \tau = -1$ and $r = a H \tau (1-\epsilon) = -1$. In slow-roll inflation, however, $\epsilon$ is small but nonzero; the energy of the inflaton field $\phi$ is dominated by a nearly flat potential $V(\phi)$. The slow-roll approximation implies $\dot{\phi} \approx -V_{,\phi} / 3H \approx (-V_{,\phi}/V) \sqrt{V/3}$, and $\epsilon \approx \frac{1}{2} (\frac{-V_{,\phi}}{V})^2 \ll 1$ which is assumed to vary only slowly. Therefore
\begin{equation}
\log a = \int H \, dt \approx \int \sqrt{\frac{V}{3}} \, \frac{d\phi}{(\frac{-V_{,\phi}}{V}) \sqrt{V/3}} \approx \big( \tfrac{V}{-V_{,\phi}} \big) \int d\phi ,
\end{equation}
hence $a \approx e^{(-V/V_{,\phi}) \phi}$ and
\begin{equation}
\tau = \int \frac{dt}{a} \approx \int \frac{e^{(V/V_{,\phi}) \phi} \, d\phi}{(\frac{-V_{,\phi}}{V}) \sqrt{V/3}} \approx - \frac{e^{(V/V_{,\phi}) \phi}}{\sqrt{V/3}}  + \int e^{(V/V_{,\phi}) \phi} \Big( \frac{-V_{,\phi}}{2V} \Big) \frac{d\phi}{\sqrt{V/3}} \approx -\frac{1}{aH} + \epsilon \, \tau ,
\end{equation}
where an integration by parts is performed and the integration constant is chosen to be zero. Consequently, $p = a H \tau \approx -1 - \epsilon$ and $r = a H \tau (1-\epsilon) \approx -1 + \mathcal{O}(\epsilon^2)$, indeed close to the fixed point. Inflation ends when the potential becomes steep and the field speeds up, which lead the flow parameters to move up across the $r = 0$ ($\epsilon = 1$) line and stop producing scale invariant modes.

The fixed point at $(0,0)$ corresponds to the \emph{adiabatic ekpyrotic} model \cite{Khoury:2009my, Khoury:2011ii}. This model relies on a slow contraction phase with $H \approx H_0 + c / t$, where $H_0 < 0$ and $c \ll 1$. During the period between $t_\text{beg} \approx 1 / H_0$ and $t_\text{end} \approx c / H_0$, $H \approx H_0$ but $\dot{H} \sim -1/t^2$, hence $\epsilon$ increases rapidly as $\sim 1/t^2$. Also within this one Hubble time $a \approx$~const, hence the conformal time $\tau \sim t$. Therefore $q^2 \sim \epsilon \sim 1/\tau^2$, which guarantees a scale invariant power spectrum. The number of scale invariant modes is given by $N \approx \log |t_\text{beg} / t_\text{end}| \approx \log (1/c)$. To see that during this period the flow parameters stay near the saddle point, note that $p = a H \tau \approx H_0 t$, and $p - r = a H \tau \epsilon \approx -t \dot{H} / H \approx c / H_0 t$. Hence for $t_\text{beg} \ll t \ll t_\text{end}$ we find $0 < p \ll 1$ and $|r| \ll 1$. Note that $(p-r)$ corresponds to a growing eigenmode of eqs.~(\ref{eq:flowp}, \ref{eq:flowr}) near the saddle point $(0,0)$; it starts extremely small at $t_\text{beg} = 1 / H_0$, $(p-r)_\text{beg} \approx c \sim e^{-N} \ll 1$, and becomes $(p-r)_\text{end} \sim \mathcal{O}(1)$ at $t_\text{end} \approx c / H_0$. We have shown that the adiabatic ekpyrotic mechanism takes place near the point $(0,0)$ on the $p > 0$ side. Notably, this fixed point may also be approached from the $p < 0$ side, which corresponds to a \emph{slow expansion} \cite{Khoury:2010gw, Joyce:2011ta} that produces scale invariant perturations in a similar way.

Finally, the fixed point at $(0,1)$ corresponds to the \emph{apex} model, first discussed in \cite{Khoury:2010gw}. In this model the universe slowly transitions from expansion to contraction. But unlike in \cite{Khoury:2010gw}, here we describe the apex model as a distinct mechanism that is independent from the adiabatic ekpyrotic slow expansion. When the universe gradually stops expansion to reach an apex, the scale factor $a$ slowly comes to a halt, and the Hubble parameter $H$ has to drop from positive to zero. Therefore, near the apex we may assume that $H \approx \dot{H}_0 t$, where the constant $\dot{H}_0 < 0$ and $t$ goes from negative to $0$. During the last Hubble time when $t > t_\text{beg} = - 1 / \sqrt{-\dot{H}_0}$, the scale factor $a \approx$~const, hence $\tau \sim t$. The scale invariant perturbations are produced by $q^2 \sim \epsilon \approx -\dot{H}_0 / H^2 \sim 1/\tau^2$. The flow parameters are given by $p = a H \tau \approx \dot{H}_0 t^2$, and $r = a H \tau (1-\epsilon) \approx 1 + \dot{H}_0 t^2$. They approach the sink point $(0,1)$ asymptotically as $t \to 0^-$. It seems that, as the log time $-\log(-\tau) \to \infty$, an infinite number of scale invariant modes would be produced. Nevertheless, before reaching the point $(0,1)$ at which $\epsilon \to \infty$, our analysis based on the quadratic action (\ref{eq:S2}) should already be expected to break down due to strong coupling \cite{Baumann:2011dt}. Therefore the number of scale invariant modes should be finite. It is also important to have an asymmetry in the cosmic evolution before and after the apex, so that the scale invariant modes are not undone in the subsequent contraction phase.

We have represented all scale invariant cosmological models with $c_s = 1$ by the flow lines on the $\tilde{s} = 0$ plane in the parameter space. They correspond to solutions $a(\tau)$ of the generalized Emden-Fowler equation imposed by the requirement of a scale invariant power spectrum \cite{Baumann:2011dt},
\begin{equation}
q^2 = a \Big( \frac{a^2}{a'} \Big)' \propto \frac{1}{\tau^2} .
\end{equation}
Our diagrammatic representation provides a clear way of classifying models by the flow lines and searching for simple analytic solutions near the fixed points. The main advantage of our approach will become clear when we consider models with a time varying speed of sound. In that case the requirement of a scale invariant power spectrum alone does not suffice to fully determine the time dependence of both functions $a(y)$ and $c_s(y)$, hence it is not possible to find solutions without imposing further constraints. On the other hand, in terms of the flow parameters, eqs.~(\ref{eq:flowp}, \ref{eq:flowr}) still hold except that $\tilde{s}$ is now another parameter that may vary. To look for new types of scale invariant cosmological models, we shall generalize our analysis to the whole parameter space $(p, r, \tilde{s})$.

\section{Fixed points} \label{sec:stat}

In the general case, the cosmological models will be represented by flow lines in the parameter space $(p, r, \tilde{s})$. Since the types of cosmological models are largely determined by the positions and properties of the fixed points, we will first locate the fixed points and then analyze the flow lines around them.

\begin{figure}[t]
\centering
\includegraphics[width=0.5\textwidth]{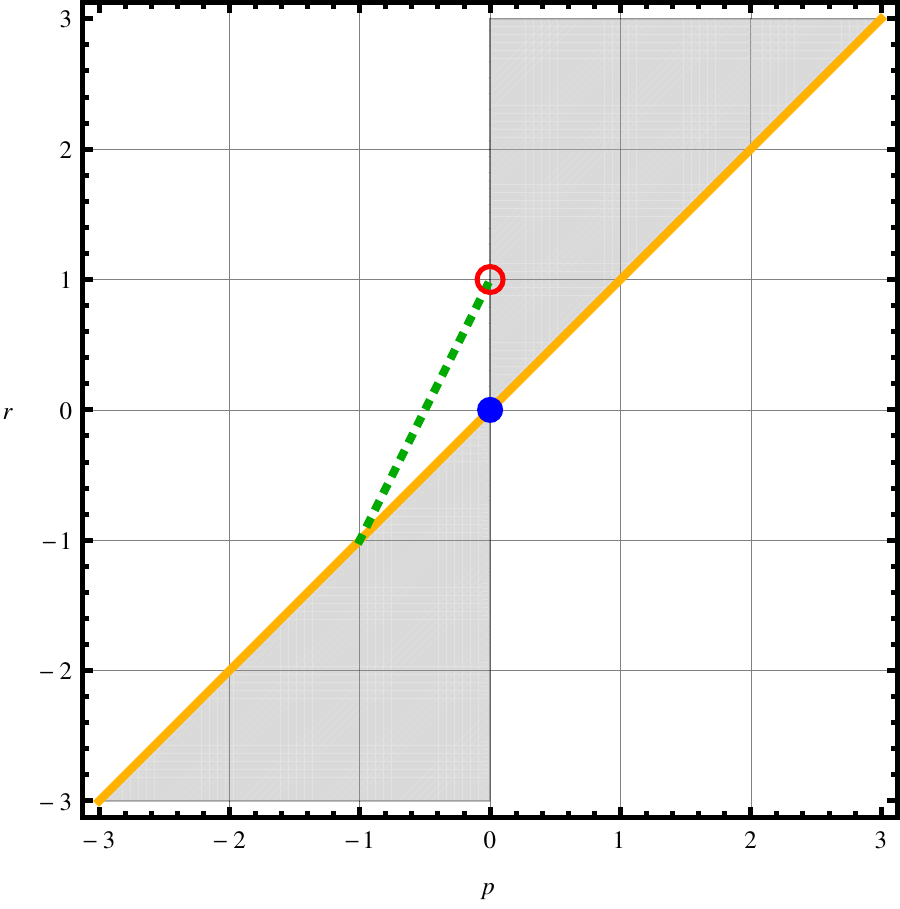}
\caption{Positions of the four fixed points in scale invariant models ($n = -1$), projected on the $p \, r$-plane for different values of $\tilde{s}$: the adiabatic ekpyrosis (solid blue circle) at $(0, 0)$ for all $\tilde{s}$, the decelerated expansion (red circle) at $(0, 1)$ for all $\tilde{s}$, the inflation/deflation point (thick orange line) at $(\tilde{s}-1, \tilde{s}-1)$, and the tachyacoustic expansion (dashed green line) at $(\frac{\tilde{s}}{2}-1, \tilde{s}-1)$ for $0 < \tilde{s} < 2$.} \label{fig:stpts}
\end{figure}

Despite the unknown dynamics of $\tilde{s}$, the fixed points of the flow equations (\ref{eq:flowp}, \ref{eq:flowr}) must satisfy
\begin{equation}
\left\{ \begin{array}{l}
- p (r - \tilde{s} + 1) = 0 , \\[4pt]
- p (2 p - 2 n - \tilde{s}) + 2 p r - r (r + 2 n + 1) = 0 .
\end{array} \right.
\end{equation}
For each value of $\tilde{s}$, there are four solutions to these equations, namely $(p, r) = (0, 0)$, $(0, - 2n - 1)$, $(\tilde{s} - 1, \tilde{s} - 1)$, and $(\tilde{s}/2 + n, \tilde{s} - 1)$. Each solution draws a fixed point in the parameter space, which traces out a line as $\tilde{s}$ varies; their positions projected on the $p \, r$-plane are shown in figure~\ref{fig:stpts} for $n = -1$.
The first three fixed points are generalizations of the ones in the $c_s = 1$ ($\tilde{s} = 0$) case to models with time varying $c_s$ ($\tilde{s} \neq 0$). They will be referred to as the \emph{adiabatic ekpyrosis} point $(0, 0)$, the \emph{decelerated expansion} point $(0, 1)$, and the \emph{inflation/deflation} point $(\tilde{s}-1, \tilde{s}-1)$. The fourth fixed point $(\tilde{s}/2-1, \tilde{s}-1)$ corresponds to the \emph{tachyacoustic expansion} scenario \cite{ArmendarizPicon:2006if, Piao:2006ja, Magueijo:2008pm, Bessada:2009ns}. It is degenerate with inflation at $\tilde{s} = 0$, but otherwise is a distinct scenario that can generate scale invariant curvature perturbations in an expanding universe with a decreasing speed of sound. Incidentally, for $n = 2$ the fourth fixed point $(\tilde{s}/2+2, \tilde{s}-1)$ corresponds to a contracting universe with a growing speed of sound described in \cite{Khoury:2008wj}.

To learn the properties of the fixed points, we shall analyze the flow lines near those points, as in the $\tilde{s} = 0$ case. In the more general case, the parameter $\tilde{s}$ may be nonzero and even time dependent. However, in the larger parameter space $(p,r,\tilde{s})$, the dynamics of the parameter $\tilde{s}$ is not constrained by the requirement of a scale invariant power spectrum that led to eqs.~(\ref{eq:flowp}, \ref{eq:flowr}). For a practical model such as a scalar field with a particular Lagrangian, the speed of sound $c_s$ is determined by the field configuration. Hence generally there is a relation between the parameters $\tilde{s}$ and $p, r$, as shown by a few examples in section~\ref{sec:exam}. Here, without being constrained by the underlying model, we will phenomenologically explore different situations in which $\tilde{s}$ can vary with respect to $p$ and $r$.

In particular, consider flow lines on a surface $\tilde{s}(p,r)$ in the parameter space that goes through one fixed point. Then near that point a small deviation in $\tilde{s}$ is given by
\begin{equation} \label{eq:deltas}
\delta \tilde{s} \approx \frac{\partial \tilde{s}}{\partial p} \, \delta p + \frac{\partial \tilde{s}}{\partial r} \, \delta r \equiv \alpha \, \delta p + \beta \, \delta r ,
\end{equation}
where $(\alpha, \beta)$ is the gradient of the function $\tilde{s}(p,r)$ at the fixed point. The flow equations (\ref{eq:flowp}, \ref{eq:flowr}) can be expanded to linear order in $\delta p, \delta r$ near that point, setting $n = -1$,
\begin{align}
- \frac{d \, \delta p}{d \log(-y)} &= - (- \alpha p + r - \tilde{s} + 1) \, \delta p - (1 - \beta) p \, \delta r , \label{eq:pertflowp} \\[4pt]
- \frac{d \, \delta r}{d \log(-y)} &= - ( (4 - \alpha) p - 2 r - \tilde{s} + 2) \, \delta p  - (- (2 + \beta) p + 2 r - 1) \, \delta r . \label{eq:pertflowr}
\end{align}
The eigenvalues to this linear system are given by
\begin{equation} \label{eq:eigs}
\left\{ \begin{array}{l}
\lambda_1 + \lambda_2 = (2 + \alpha + \beta) p - 3 r + \tilde{s} , \\[4pt]
\lambda_1 \lambda_2 = (3 \alpha + 4 \beta - 4) p^2 + ( 3 \tilde{s} + \alpha + \beta - 4 ) p - (2 \alpha + 3 \beta) p r + (2 r - 1)(r - \tilde{s} + 1) .
\end{array} \right.
\end{equation}
These eigenvalues determine whether a fixed point is a source, a sink, or a saddle point.

In the following we analyze each of the four fixed points and derive approximate solutions of the nearby flow lines. The corresponding types of cosmological models are described and verified to create scale invariant curvature perturbations. These models rely on having a manifestly time dependent speed of sound; constraints from the strong coupling problem and superluminality are discussed in section~\ref{sec:cons}.

\subsection{adiabatic ekpyrosis: $(p, r) = (0, 0)$} \label{sec:adek}

This fixed point stays in the same position for any value of $\tilde{s}$. Like in the $\tilde{s} = 0$ case, the corresponding cosmological model is either a slow contraction if $0 < p \ll 1$ or a slow expansion if $-1 \ll p < 0$. The eigenvalues from eqs.~(\ref{eq:eigs}) can be explicitly found to be $\lambda_1 = 1$, $\lambda_2 = \tilde{s} - 1$. Because the first eigenvalue is positive, this fixed point can be either a saddle point or a source of flow depending on the value of the second eigenvalue. Specifically, it is a saddle point if $\tilde{s} < 1$, which is qualitatively the same as in the case $\tilde{s} = 0$ discussed in section~\ref{sec:flow}. However, for $\tilde{s} > 1$ we find a new situation where this point behaves like a source.

In this new situation with $\tilde{s} > 1$, the flow lines emerge from the point $(0,0)$ in the asymptotic past $y \to -\infty$. In that limit, since $p, r \to 0$, both $a$ and $H$ approach constant values, $a \approx a_{-\infty}$ and $H \approx H_{-\infty}$. Take $\alpha = \beta = 0$ for simplicity, so that $c_s \sim (-y)^{\tilde{s}}$. Then in the asymptotic past,
\begin{equation} \label{eq:physt}
t = \int \frac{a}{c_s} dy \sim \int_{-\infty}^{y} \frac{a_{-\infty}}{(-y)^{\tilde{s}}} dy \sim \frac{1}{(-y)^{\tilde{s}-1}} ,
\end{equation}
which goes from $0$ to $\infty$. That is, the asymptotic past corresponds to finite physical time $t$.

To see the cosmic evolution in this situation, it is better to solve the flow equations and express $a$ and $H$ in terms of $t$. Near $(p,r) = (0,0)$, the eigenmodes of eqs.~(\ref{eq:pertflowp}, \ref{eq:pertflowr}) are given by $p \propto 1 / (-y)^{\tilde{s}-1}$ and $r - p \propto 1 / (-y)$. Using the definitions of $p$ and $r$, one finds
\begin{align}
& \log \Big( \frac{a}{a_{-\infty}} \Big) \sim \frac{1}{(-y)^{\tilde{s}-1}} \sim t , \\[4pt]
& \log \Big( \frac{H}{H_{-\infty}} \Big) \sim \frac{1}{(-y)} \sim t^{1/(\tilde{s}-1)} .
\end{align}
Therefore, in the limit $t \to 0$,
\begin{align}
a &\approx a_{-\infty} (1 + H_{-\infty} t) , \\[4pt]
H &\approx H_{-\infty} \Big( 1 - C (H_{-\infty} t)^{1/(\tilde{s}-1)} \Big) , \\[4pt]
c_s &\sim (-y)^{\tilde{s}} \sim t^{-\tilde{s} / (\tilde{s}-1)} ,
\end{align}
where we assume $C \sim \mathcal{O}(1) > 0$. For $t \ll 1/|H_{-\infty}|$ it is approximately true that $a \approx a_{-\infty}$ and $H \approx H_{-\infty}$. Under this approximation,
\begin{equation}
\epsilon \approx \frac{-\dot{H}}{H_{-\infty}^2} \approx \frac{C}{\tilde{s}-1} (H_{-\infty} t)^{(2-\tilde{s}) / (\tilde{s} - 1)} \sim (-y)^{\tilde{s}-2} ,
\end{equation}
which guarantees the condition $q^2 \sim \epsilon / c_s \sim 1 / (-y)^2$ for generating scale invariant curvature perturbations.

A particularly interesting example is when $\tilde{s} = 2$. This corresponds to a cosmological model in which $H \approx H_{-\infty} - C H_{-\infty}^2 t$. For $0 < t \ll 1 / |H_{-\infty}|$, it gives $a \approx a_{-\infty}$ and $\epsilon \approx C =$~const. If the speed of sound satisfies $c_s \sim 1/t^2$, then one finds $(-y) \sim 1/t$, hence $q^2 \sim 1 / c_s \sim 1 / (-y)^2$. That is, the scale invariance of the power spectrum is achieved by having a uniquely time dependent speed of sound. This model can be considered as dual to either inflation or adiabatic ekpyrosis in which the $1/(-y)^2$ dependence of $q^2 = a^2 \epsilon / c_s$ comes solely from either $a$ or $\epsilon$. Like in adiabatic ekpyrosis, this mechanism works in either a contracting or an expanding universe according to whether $H_{-\infty}$ is negative or positive.

\subsection{decelerated expansion: $(p, r) = (0, 1)$} \label{sec:apex}

This fixed point represents a decelerated cosmic expansion, since near this point $p < 0$ and $\epsilon = (p-r)/p \gg 1$. The eigenvalues from eqs.~(\ref{eq:eigs}) are $\lambda_1 = -1$, $\lambda_2 = \tilde{s} - 2$, which again do not depend on $\alpha, \beta$. For $\tilde{s} < 2$ this fixed point is a sink, the same as the apex model in the $\tilde{s} = 0$ case. But for $\tilde{s} > 2$ it becomes a saddle point, which gives rise to a new scenario.

In this new scenario with $\tilde{s} > 2$, the flow lines do not asymptote towards $p = 0$ but move away from it instead. Hence the universe does not approach an apex. Indeed, near the point $(0,1)$, $a \approx$~const and $H \sim (-y)$; but unlike in the apex model where $H \to 0$ as $y \to 0^-$, here the flow line passes by the saddle point while $y$ remains finite. The unstable eigenmode of eqs.~(\ref{eq:pertflowp}, \ref{eq:pertflowr}) is given by $p \propto -1 / (-y)^{\tilde{s}-2}$, which implies $\epsilon \approx -1 / p \propto (-y)^{\tilde{s}-2}$. For $\tilde{s} > 2$, $\epsilon$ decreases with time rather than increases.

To see the cosmic evolution in physical time $t$, assume again $\alpha = \beta = 0$ so that $c_s \sim (-y)^{\tilde{s}}$. Then for $a \approx$~const one obtains $t \sim 1 / (-y)^{\tilde{s}-1}$ as in eq.~(\ref{eq:physt}), where $t > 0$ and increases. Thus the cosmological model can be described by a decelerated expansion with
\begin{equation} \label{eq:Hdecel}
H \sim t^{-1/(\tilde{s}-1)} , \quad c_s \sim t^{-\tilde{s}/(\tilde{s}-1)} .
\end{equation}
We can check that $\dot{H} \sim - t^{-\tilde{s}/(\tilde{s}-1)}$ and hence $\epsilon \sim t^{-(\tilde{s}-2) / (\tilde{s}-1)} \sim (-y)^{\tilde{s}-2}$, ensuring $q^2 \sim \epsilon / c_s \sim 1/(-y)^2$. To be consistent, since by eq.~(\ref{eq:Hdecel}),
\begin{equation}
\log \Big( \frac{a}{a_0} \Big) \sim t^{(\tilde{s}-2) / (\tilde{s}-1)} ,
\end{equation}
the approximation $a \approx$~const holds for sufficiently small $t$. This approximation breaks down at finite time before $H$ reaches $0$, showing again that it is different from an apex model.

There is a marginal case where $\tilde{s} = 2 + 2 p_0$ with $-1 \ll p_0 < 0$. In that case $p \approx p_0 =$~const, hence $a \sim (-y)^{p_0}$. Since $c_s \sim (-y)^{2+2p_0}$, the physical time $t \sim 1/(-y)^{1+p_0} > 0$. The cosmic evolution is given by
\begin{equation}
a \sim t^{-p_0/(1+p_0)} , \quad H \approx \frac{(-p_0)}{(1+p_0) t} , \quad c_s \sim \frac{1}{t^2} .
\end{equation}
This is a slow expansion with constant $\epsilon = - (1+p_0) / p_0 \gg 1$. Therefore $q^2 \sim a^2 / c_s \sim 1/(-y)^2$, which produces scale invariant curvature perturbations. This model describes an extremely decelerated expansion with a rapidly decreasing speed of sound, which is the $\epsilon \gg 1$ limit of the tachyacoustic model to be discussed in section~\ref{sec:tach}. Interestingly, this model has a contraction analog with constant $\epsilon \gg 1$, represented by a flow line in the $p > 0$ half plane. This new ``acoustic ekpyrotic'' model will be discussed in section~\ref{sec:eps-cs}.

\subsection{inflation/deflation: $(p, r) = (\tilde{s} - 1, \tilde{s} - 1)$} \label{sec:inf}

This fixed point describes a de Sitter universe with $\epsilon = (p-r)/p = 0$. The position of this fixed point depends on $\tilde{s}$. For $\tilde{s} < 1$ it lies on the border $p = r < 0$ in the left half plane, the same as inflation; but for $\tilde{s} > 1$ it moves to the right half plane with $p = r > 0$ (the case $\tilde{s} = 0$ coincides with adiabatic ekpyrosis). The eigenvalues from eqs.~(\ref{eq:eigs}) are $\lambda_1 = \tilde{s}$, $\lambda_2 = (\alpha + \beta - 1) (\tilde{s} - 1)$. Depending on the value of $\alpha + \beta$, this point can be a saddle, a source, or a sink, as listed in table~\ref{tab:stpts}.

Here we consider the new situation when $\tilde{s} > 1$ and $p = r > 0$, which corresponds to an extremely rapid cosmic contraction, or ``deflation''. Specifically, near this fixed point, $p \approx \tilde{s}-1$ and $r-p \approx 0$, hence $a \sim (-y)^{\tilde{s}-1}$ and $H \approx$~const. The eigenmode that corresponds to the first eigenvalue is $(p-r) \propto 1/(-y)^{\tilde{s}}$. Therefore $\epsilon \approx (p-r) / (\tilde{s}-1) \sim 1/(-y)^{\tilde{s}}$, which is crucial to ensure $q^2 = a^2 \epsilon / c_s \sim 1/(-y)^2$. Now that $\epsilon$ is rapidly increasing, it is clear that the flow parameters can only stay near the deflation point for a finite time.

For $a \sim (-y)^{\tilde{s}-1}$ and $c_s \sim (-y)^{\tilde{s}}$, the physical time $t$ is given by
\begin{equation}
t = \int \frac{a}{c_s} dy \sim \int \frac{(-y)^{\tilde{s}-1}}{(-y)^{\tilde{s}}} dy = - \log (-y) ,
\end{equation}
i.e. proportional to the number of scale invariant modes. In physical time $t$, if $H \approx H_0 < 0$, then $a \sim e^{H_0 t}$. Comparing that to $a \sim (-y)^{\tilde{s}-1}$, one finds
\begin{equation}
t \approx \frac{\tilde{s}-1}{H_0} \log(-y) , \quad \mbox{or} \quad (-y) \approx e^{H_0 t / (\tilde{s}-1)} .
\end{equation}
The crucial part of this model is to have a rapidly growing $\epsilon$ and a rapidly decreasing $c_s$,
\begin{equation}
\epsilon \approx \frac{\epsilon_0}{(-y)^{\tilde{s}}} \approx \epsilon_0 \, e^{- \frac{\tilde{s}}{\tilde{s}-1} H_0 t} , \quad c_s \sim (-y)^{\tilde{s}} \sim e^{\frac{\tilde{s}}{\tilde{s}-1} H_0 t} .
\end{equation}
The de Sitter approximation $\epsilon \ll 1$ is valid until $t_\text{end} \approx \frac{(\tilde{s}-1)/\tilde{s}}{(-H_0)} \log (\frac{1}{\epsilon_0})$. If we assume that $\epsilon_0 \ll 1$, then there can be many numbers of Hubble times before $t_\text{end}$.

Therefore, this deflation model is described by a rapid cosmic contraction with
\begin{equation} \label{eq:Hdef}
H \approx H_0 \Big( 1 + \tfrac{\tilde{s}-1}{\tilde{s}} \, \epsilon_0 \, e^{- \frac{\tilde{s}}{\tilde{s}-1} H_0 t} \Big) ,
\end{equation}
which satisfies $H \approx H_0$ for $t \ll t_\text{end}$. The scale invariance of the power spectrum is achieved by an interplay between
\begin{equation} \label{eq:aecdef}
a \sim e^{H_0 t} , \quad \epsilon \approx \epsilon_0 \, e^{- \frac{\tilde{s}}{\tilde{s}-1} H_0 t} , \quad c_s \sim e^{\frac{\tilde{s}}{\tilde{s}-1} H_0 t} .
\end{equation}

So far the dependence of the model on $\alpha, \beta$ has been neglected as we implicitly assumed a nearly constant $\tilde{s}$. For $\alpha + \beta > 1$ the fixed point is actually a source where the flow line emerges in the asymptotic past, so the above description works even better at early times $t \to -\infty$. However, for $\alpha + \beta < 1$ the fixed point is a saddle point instead. In that case, when extrapolating backwards in time, the flow parameters would deviate from the saddle point at certain time $t_\text{beg}$. This time is controlled by the second eigenvalue, $t_\text{beg} \approx \frac{1}{(1 - \alpha - \beta) H_0} \log (\frac{1}{\epsilon_0})$. The model described above is valid between $t_\text{beg}$ and $t_\text{end}$ to produce roughly $N \sim \log (\frac{1}{\epsilon_0})$ scale invariant modes.

Note that eqs.~(\ref{eq:Hdef}, \ref{eq:aecdef}) also apply for $\tilde{s} < 1$ and $H_0 > 0$, in which case they describe an exponentially expanding universe with a time varying speed of sound, or ``acoustic inflation''. This model is less interesting since inflation with $\epsilon \ll 1$ and a constant $c_s$ can already produce scale invariant perturbations. In contrast, the deflation model relies on having a rapidly decreasing $c_s$ as well as a small but increasing $\epsilon$.

\subsection{tachyacoustic expansion: $(p, r) = (\tilde{s}/2 - 1, \tilde{s} - 1)$} \label{sec:tach}

The position of this fixed point also depends on $\tilde{s}$. Moreover, since $\epsilon = (p-r)/p = \tilde{s} / (2-\tilde{s})$, the null energy condition $\epsilon \geq 0$ requires $0 < \tilde{s} < 2$ (the cases $\tilde{s} = 0, 2$ coincide with inflation and apex respectively). Thus in the $p \, r$-plane, this fixed point lies on the line segment $r = 2p + 1$ for $-1 < p < 0$, as shown in figure~\ref{fig:stpts}. Since $p < 0$ and $\tilde{s} > 0$, it describes an expanding universe with a decreasing speed of sound.

Here we show that this model precisely corresponds to the \emph{tachyacoustic expansion} scenario \cite{ArmendarizPicon:2006if, Piao:2006ja, Magueijo:2008pm, Bessada:2009ns}. Specifically, at the point $(p, r) = (\tilde{s}/2-1, \tilde{s}-1)$ for certain value of $\tilde{s}$, the model is described by $a \sim (-y)^{\tilde{s}/2-1}$, $H \sim (-y)^{\tilde{s}/2}$, and $c_s \sim (-y)^{\tilde{s}}$. From eq.~(\ref{eq:n=}) one finds $\tilde{\eta} = 0$, consistent with $\epsilon = \tilde{s} / (2-\tilde{s}) =$~const. Therefore $q^2 \sim a^2 / c_s \sim 1/(-y)^2$, ensuring a scale invariant power spectrum.

In this model, the physical time $t$ is given by
\begin{equation}
t = \int \frac{a}{c_s} dy \sim \int \frac{dy}{(-y)^{\tilde{s}/2+1}} \sim \frac{1}{(-y)^{\tilde{s}/2}} .
\end{equation}
Thus in physical time $t$, $a \sim t^{(2-\tilde{s})/\tilde{s}}$, $H = (2-\tilde{s}) / (\tilde{s} \, t)$, and $c_s \sim 1/t^2$. We can check that $\epsilon = -\dot{H}/H^2 = \tilde{s} / (2-\tilde{s})$ as expected. In terms of the parameter $s \equiv \dot{c_s} / H c_s = \tilde{s} / p$, this relation can be expressed as $s = -2 \epsilon$, which agrees perfectly with the tachyacoustic scenario \cite{ArmendarizPicon:2006if, Piao:2006ja, Magueijo:2008pm, Bessada:2009ns, Khoury:2008wj}. This scenario is emphatically different from inflation since here $\epsilon$ need not be small at all. In particular, the limit $\epsilon \gg 1$ corresponds precisely to the marginal case in section~\ref{sec:apex}.

For flow lines near this fixed point, the eigenvalues in eq.~(\ref{eq:eigs}) become
\begin{equation} \label{eq:tac-eigs}
\left\{ \begin{array}{l}
\lambda_1 + \lambda_2 = \big( \frac{\alpha + \beta}{2} - 1 \big) \tilde{s} - (\alpha + \beta - 1) , \\[4pt]
\lambda_1 \lambda_2 = \frac{1}{2} \big( \frac{\alpha}{2} + \beta - 1 \big) \tilde{s} (2 - \tilde{s}) .
\end{array} \right.
\end{equation}
Therefore this point is a saddle point if $\frac{\alpha}{2} + \beta < 1$, otherwise a source or a sink depending on whether $\big( \frac{\alpha + \beta}{2} - 1 \big) \tilde{s} - (\alpha + \beta - 1) > 0$ or $< 0$, as listed in table~\ref{tab:stpts}.

\begin{table}[h]
\centering
\renewcommand{\arraystretch}{1.5}
\begin{tabular}{|c|c|c|c|c|}
\hline
fixed  & adia ekpy & decel exp & inf / def                     & tachyacoustic \\[-6pt]
points      & $(0,0)$   & $(0,1)$   & $(\tilde{s}-1, \tilde{s}-1)$  & $(\frac{\tilde{s}}{2} -1, \tilde{s}-1)$ \\
\hline
\multirow{2}{*}{source} & \multirow{2}{*}{$\tilde{s} > 1$}      & \multirow{2}{*}{$\diagup$}            & $0 < \tilde{s} < 1, \alpha + \beta < 1$   & $\frac{\alpha}{2} + \beta > 1,$ \\
\cline{4-4}
&   &   & $\tilde{s} > 1, \alpha + \beta > 1$ & $\big( \frac{\alpha + \beta}{2} - 1 \big) \tilde{s} > (\alpha + \beta - 1)$   \\
\hline
\multirow{3}{*}{saddle} & \multirow{3}{*}{$\tilde{s} \leq 1$}   & \multirow{3}{*}{$\tilde{s} > 2$}  & $\tilde{s} \leq 0, \alpha + \beta < 1$    & \multirow{3}{*}{$\frac{\alpha}{2} + \beta < 1$}    \\
\cline{4-4}
&   &   & $0 < \tilde{s} < 1, \alpha + \beta > 1$   &   \\
\cline{4-4}
&   &   & $\tilde{s} > 1, \alpha + \beta < 1$   &   \\
\hline
\multirow{2}{*}{sink}   & \multirow{2}{*}{$\diagup$}    & \multirow{2}{*}{$\tilde{s} \leq 2$}   & \multirow{2}{*}{$\tilde{s} \leq 0, \alpha + \beta > 1$}   & $\frac{\alpha}{2} + \beta > 1,$ \\
&   &   &   & $\big( \frac{\alpha + \beta}{2} - 1 \big) \tilde{s} < (\alpha + \beta - 1)$   \\
\hline
\end{tabular}
\caption{Positions and properties of the fixed points for flow lines on a constrained surface $\tilde{s}(p,r)$ with local gradient $(\alpha,\beta)$.} \label{tab:stpts}
\end{table}

\section{Examples} \label{sec:exam}
\addtocontents{toc}{\protect\setcounter{tocdepth}{1}}

We have analyzed the flow lines near the fixed points under certain constraint function $\tilde{s}(p,r)$ with local gradient $(\alpha, \beta)$. For concreteness, consider examples that are motivated by scalar field models. Suppose the universe is dominated by a scalar field $\phi$ with Lagrangian $\mathcal{L} = P(\phi,X)$, where $X \equiv -\frac{1}{2} (\partial \phi)^2$. It is shown in appendix~\ref{sec:A} that, for a given function $P(\phi,X)$, the parameter $\eta \equiv \dot{\epsilon} / H \epsilon$ can be locally determined as a function of $\epsilon$ and $s$. Such a function $\eta(\epsilon,s)$ suffices to provide the constraint between the parameters $p, r$ and $\tilde{s}$.

\subsection{example I: $\eta = m \, s$} \label{sec:eps-cs}

Consider first the \textit{k}-essence like model in which the Lagrangian is factorizable as $\mathcal{L} = F(\phi) \bar{P}(X)$ \cite{ArmendarizPicon:2000ah}. In this case both the equation of state $w$ and the speed of sound $c_s$ depend only on $X$, namely
\begin{equation}
w = \frac{\bar{P}}{2X \bar{P}_{,X} - \bar{P}} \, , \quad c_s^2 = \frac{\bar{P}_{,X}}{2X \bar{P}_{,XX} + \bar{P}_{,X}} \, .
\end{equation}
Therefore the parameter $\epsilon = \frac{3}{2}(1+w)$ is locally a function of $c_s$, which is determined by the form of the function $\bar{P}(X)$.

As a simple example, take the function $\epsilon(c_s)$ to be a power law, $\epsilon \propto {c_s}^m$, with a constant exponent $m$. Such a relation has been considered in different models. The case with $\epsilon \propto c_s^2$ ($m = 2$) appears in nonsingular bouncing models where a ghost condensate field violates the null energy condition. The case with $\epsilon \propto c_s$ ($m = 1$) is considered in \cite{Joyce:2011kh} for which some contribution to the non-Gaussianity is scale invariant, so as to avoid the strong coupling problem. And the case with $\epsilon =$~const ($m = 0$) leads to the tachyacoustic model.

This above relation $\epsilon \propto {c_s}^m$ is equivalent to having $\eta = m \, s$, a special case in which $\eta$ only depends on $s$. In terms of the flow parameters, it can be expressed as $\tilde{\eta} = m \, \tilde{s}$. Then eq.~(\ref{eq:n=}) implies, for $m \neq 1$,
\begin{equation}
\tilde{s} = \frac{2}{1-m} \, p + \frac{2 n}{m-1} .
\end{equation}
For a constant value of $n$, this function $\tilde{s}(p,r)$ determines a plane in the $(p,r,\tilde{s})$ parameter space that is parallel to the $r$-axis and goes through the point $(p, \tilde{s}) = (n, 0)$. Comparing it to eq.~(\ref{eq:deltas}), we identify $\alpha = \frac{2}{1-m}$, $\beta = 0$.

To find the fixed points and their properties in this example, we follow the analysis in section~\ref{sec:stat} and take $n = -1$. The adiabatic ekpyrotic fixed point is at $(p, r, \tilde{s}) = (0, 0, \frac{2}{1-m})$, which is a source if $\tilde{s} = \frac{2}{1-m} > 1$, i.e. $-1 < m < 1$, or a saddle point otherwise. The decelerated expansion point is at $(0, 1, \frac{2}{1-m})$, which is a saddle point if $\tilde{s} = \frac{2}{1-m} > 2$, i.e. $0 < m < 1$, or a sink otherwise. The inflation point is at $(-1, -1, 0)$ regardless of $m$; since $\tilde{s} = 0$, it is a sink if $\alpha + \beta = \frac{2}{1-m} > 1$, i.e. $-1 < m < 1$, or a saddle point otherwise. For $m \neq 0$, the tachyacoustic point is also at $(-1, -1, 0)$, degenerate with inflation; but for $m = 0$, it can be any point $(\tilde{s}/2 - 1, \tilde{s} - 1, \tilde{s})$ with $0 < \tilde{s} < 2$ --- this is just the tachyacoustic model discussed in section~\ref{sec:tach}. These fixed points are summarized in table~\ref{tab:epsilon-cs}.

\begin{table}[h!]
\centering
\renewcommand{\arraystretch}{1.5}
\begin{tabular}{|c|c|c|c|c|}
\hline
$m$                                     & adia ekpy                   & decel exp                      & inflation             & tachyacoustic \\[-6pt]
($\alpha = \frac{2}{1-m}, \beta = 0$)   & $(0, 0, \frac{2}{1-m})$   & $(0, 1, \frac{2}{1-m})$   & $(-1,-1,0)$           & $(\frac{\tilde{s}}{2} - 1, \tilde{s} - 1, \tilde{s}), 0 < \tilde{s} < 2$ \\
\hline
$m \leq -1$                             & saddle                    & \multirow{3}{*}{sink}     & saddle                & \multirow{2}{*}{$\diagup$} \\
\cline{1-2} \cline{4-4}
$-1 < m < 0$                            & \multirow{3}{*}{source}   &                           & \multirow{4}{*}{sink} & \\
\cline{1-1} \cline{5-5}
$m = 0$                                 &                           &                           &                       & sink \\
\cline{1-1} \cline{3-3} \cline{5-5}
$0 < m < 1$                             &                           & saddle                    &                       & \multirow{3}{*}{$\diagup$} \\
\cline{1-3}
$m = 1$                                 & $\diagup$                 & $\diagup$                 &                       & \\
\cline{1-4}
$m > 1$                                 & saddle                    & sink                      & saddle                & \\
\hline
\end{tabular}
\caption{The fixed points in the model $\epsilon \propto {c_s}^m$ ($\eta = m \, s$).} \label{tab:epsilon-cs}
\end{table}

To be complete, the $m = 1$ case is treated separately here. Since $\tilde{\eta} = \tilde{s}$, from eq.~(\ref{eq:n=}) one finds $p = -1$, then eq.~(\ref{eq:flowp}) implies $r = \tilde{s} - 1$. The null energy condition $\epsilon = 1 + r \geq 0$ further restricts the parameter space to a half line $\{ p = -1, \, r = \tilde{s} - 1 \geq -1 \}$. In this restricted parameter space, neither the adiabatic ekpyrosis nor the decelerated expansion exists, whereas inflation can still happen at $(p, r, \tilde{s}) = (-1, -1, 0)$ (degenerate with the tachyacoustic point). This point is a sink which the flow parameters quickly approach, as can be found by solving eq.~(\ref{eq:flowr}). This model is described by $a \propto 1/(-y)$ and $c_s \propto \epsilon = -1 / \log(y/y_0)$, where $y_0 < y < 0$, as found in \cite{Joyce:2011kh}.

\begin{figure}[t]
\centering
\includegraphics[width=3.5in]{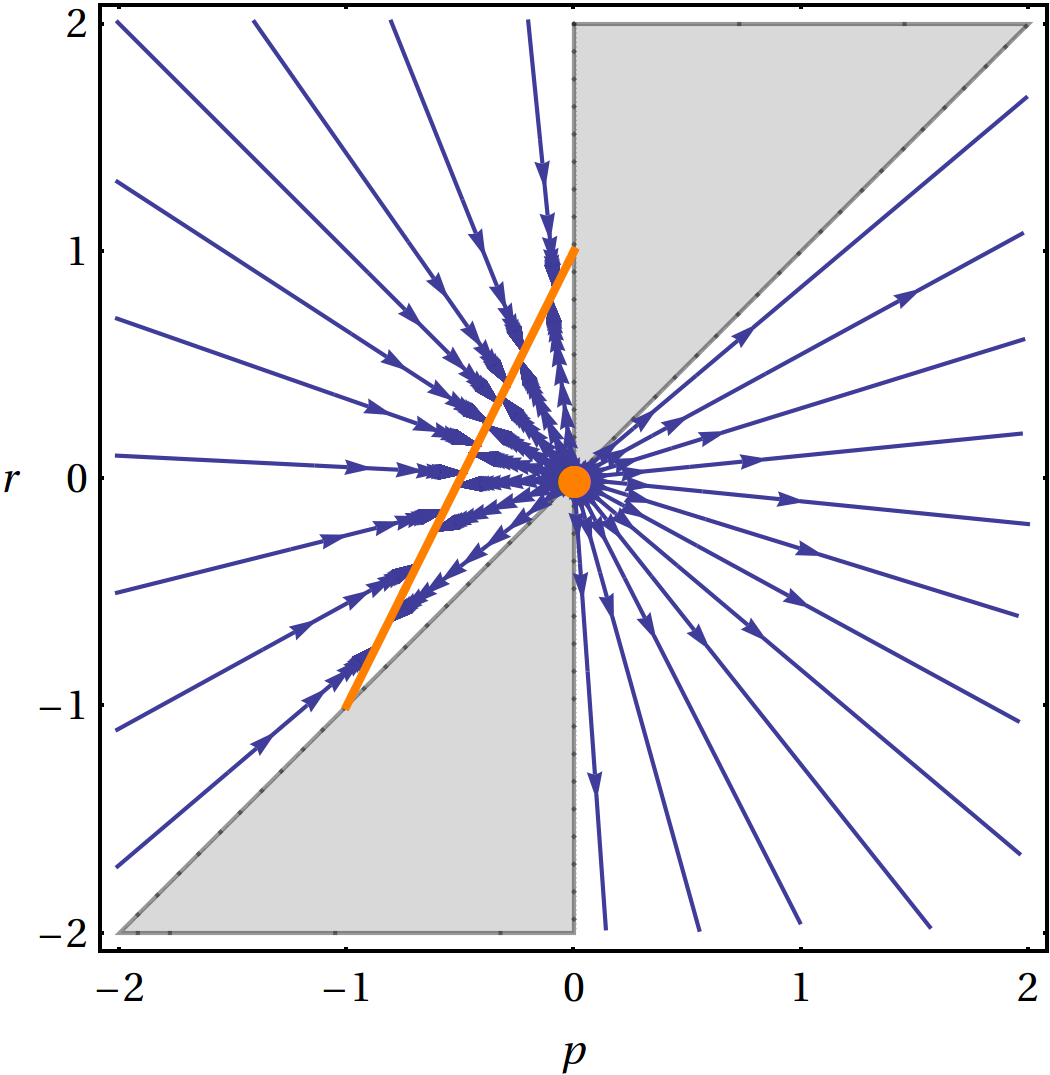}
\caption{Flow lines for models with $\epsilon =$~const. The adiabatic ekpyrosis point forms a source at the origin, and the tachyacoustic expansion points form an attractor line connecting the decelerated expansion point and the inflation point. In this and the following figures each line segment between adjacent arrowheads represents $0.5$ e-fold of scale invariant modes.} \label{fig:m=0}
\end{figure}

Here we present in detail the case with $m = 0$. The parameter space is constrained to the plane $\tilde{s} = 2 p + 2$. The flow lines have radial trajectories, as shown in figure~\ref{fig:m=0}.
It is clear that the adiabatic ekpyrosis point at the origin is a source, whereas the tachyacoustic points form an attractor line segment, with the decelerated expansion and the inflation points at the two ends. Since the flow lines in the left half plane quickly approach the attractor points, the corresponding cosmological models can be effectively described by the tachyacoustic expansion in section~\ref{sec:tach}. On the other hand, the flow lines emerging from the origin in the right half plane correspond to contraction models with a decreasing speed of sound. In particular, the trajectories near the lower half $r$-axis have $\epsilon = \mbox{const} \gg 1$ --- they represent precisely the acoustic ekpyrotic contraction mentioned in section~\ref{sec:apex}.

To describe this ``acoustic ekpyrotic'' model, we analytically solve the condition $q^2 = a^2 \epsilon / c_s \propto 1/(-y)^2$ with the constraint
\begin{equation}
\epsilon = \frac{c_s}{a} \Big( \frac{a^2}{c_s \, a'} \Big)^\prime = \mbox{const} .
\end{equation}
The result is
\begin{equation}
a(y) \propto \bigg| \frac{1}{(-y_c)} - \frac{1}{(-y)} \bigg|^{\frac{1}{\epsilon+1}} ,
\end{equation}
where $y < y_c < 0$; it describes a contracting universe that crunches at $y_c$. Accordingly, the flow parameters can be found as
\begin{align}
p &\equiv \frac{d \log a}{d \log (-y)} = \frac{1}{\epsilon + 1} \, \frac{1}{\big( \frac{y}{y_c} \big) - 1} , \\
r &= (1 - \epsilon) p = - \frac{\epsilon - 1}{\epsilon + 1} \, \frac{1}{\big( \frac{y}{y_c} \big) - 1} .
\end{align}
(Incidentally, $y < 0 < y_c$ would describe the flow lines between the origin and the attractor line, and $y_c < y < 0$ the flow lines to the left of the attractor line.) Here these parameters vary with time as they flow along the trajectories, unlike near the fixed points where they lead to simple power-laws.

Since $c_s \propto a^2 (-y)^2$, the physical time $t$ is given by
\begin{equation}
t \propto \int \frac{a \, dy}{a^2 (-y)^2} \propto - \bigg[ \frac{1}{(-y_c)} - \frac{1}{(-y)} \bigg]^{\frac{\epsilon}{\epsilon+1}} ,
\end{equation}
which goes from $t_0 \propto -(-y_c)^{-\epsilon/(\epsilon+1)}$ to $0^-$ as $y$ goes from $-\infty$ to $y_c$. From this we recover
\begin{equation}
a \propto (-t)^{1/\epsilon} , \quad H = \frac{1}{\epsilon \, t} ,
\end{equation}
which exactly describes an ekpyrotic contraction for $\epsilon \gg 1$. The scale invariance of the power spectrum relies on a decreasing speed of sound,
\begin{equation}
c_s \propto \frac{(-t)^{2/\epsilon}}{\Big( 1 - \big( \frac{t}{t_0} \big)^{1+1/\epsilon} \Big)^2} \approx \frac{1}{\Big( 1 - \big( \frac{t}{t_0} \big) \Big)^2} ,
\end{equation}
where the approximation is due to $a \approx$~const during the ekpyrotic phase. Note that as $y \to -\infty$, or $t \to t_0$, both $a$ and $H$ are finite. Near that limit the flow parameters asymptotically emerge from the adiabatic ekpyrotic point, which is the situation described in section~\ref{sec:adek}; hence in that limit the exact solutions above can be approximated by the results there.

\subsection{example II: $\eta = a \, \epsilon + b$} \label{sec:eta-eps}

Another example is motivated by a particular $\phi$ dependence of the function $F(\phi)$ in the \textit{k}-essence model, namely $F(\phi) \propto 1/\phi^2$ \cite{ArmendarizPicon:2000dh, ArmendarizPicon:2000ah}. In this case, it is shown in appendix~\ref{sec:A} that the parameter $\eta$ can be determined as a function of $\epsilon$ only.

For simplicity, take $\eta(\epsilon)$ to be a linear function, $\eta = c \, \epsilon + b$, with some constants $b, c$. Since by definition $\eta \equiv \dot{\epsilon} / H \epsilon = d \log \epsilon / d \log a$, such a function $\eta(\epsilon)$ can be used to study the running of the parameter $\epsilon$. Namely, in an expanding universe if $c < 0$ then $\epsilon$ should approach the fixed point $\epsilon_* = -b/c$, whereas in a contracting universe the fixed point is an attractor if $c > 0$.

Using eqs.~(\ref{eq:epsilon}) and (\ref{eq:n=}), the relation $\eta = c \, \epsilon + b$ can be expressed as
\begin{equation}
\tilde{s} = (b + c + 2) \, p - c \, r - 2 n ,
\end{equation}
which also determines a plane in the parameter space $(p, r, \tilde{s})$ for constant values of $n$. In comparison with (\ref{eq:deltas}) we identify $\alpha = b + c + 2$, $\beta = - c$.

Following the general analysis, we look for fixed points with $n = -1$. The adiabatic ekpyrotic point is at $(0, 0, 2)$, which is a source since $\tilde{s} > 1$. The decelerated expansion point is at $(0, 1, 2-c)$, which is a sink if $\tilde{s} = 2 - c < 2$, i.e. $c > 0$, or a saddle point otherwise. The inflation/deflation point is at $(\frac{-1}{b+1}, \frac{-1}{b+1}, \frac{b}{b+1})$ for $b \neq -1$, which is a sink if $\tilde{s} = \frac{b}{b+1} < 0$ and $\alpha + \beta = b + 2 > 1$, i.e. $-1 < b < 0$, or a saddle point otherwise. The tachyacoustic point is at $(\frac{c}{b-c}, \frac{b+c}{b-c}, \frac{2b}{b-c})$ provided $0 < \tilde{s} = \frac{2b}{b-c} < 2$; it is a saddle point if $b < 0 < c$, or a sink if $c < 0 < b$. These cases are listed in table~\ref{tab:eta-epsilon}.

\begin{figure}[t]
\centering
\includegraphics[width=3.5in]{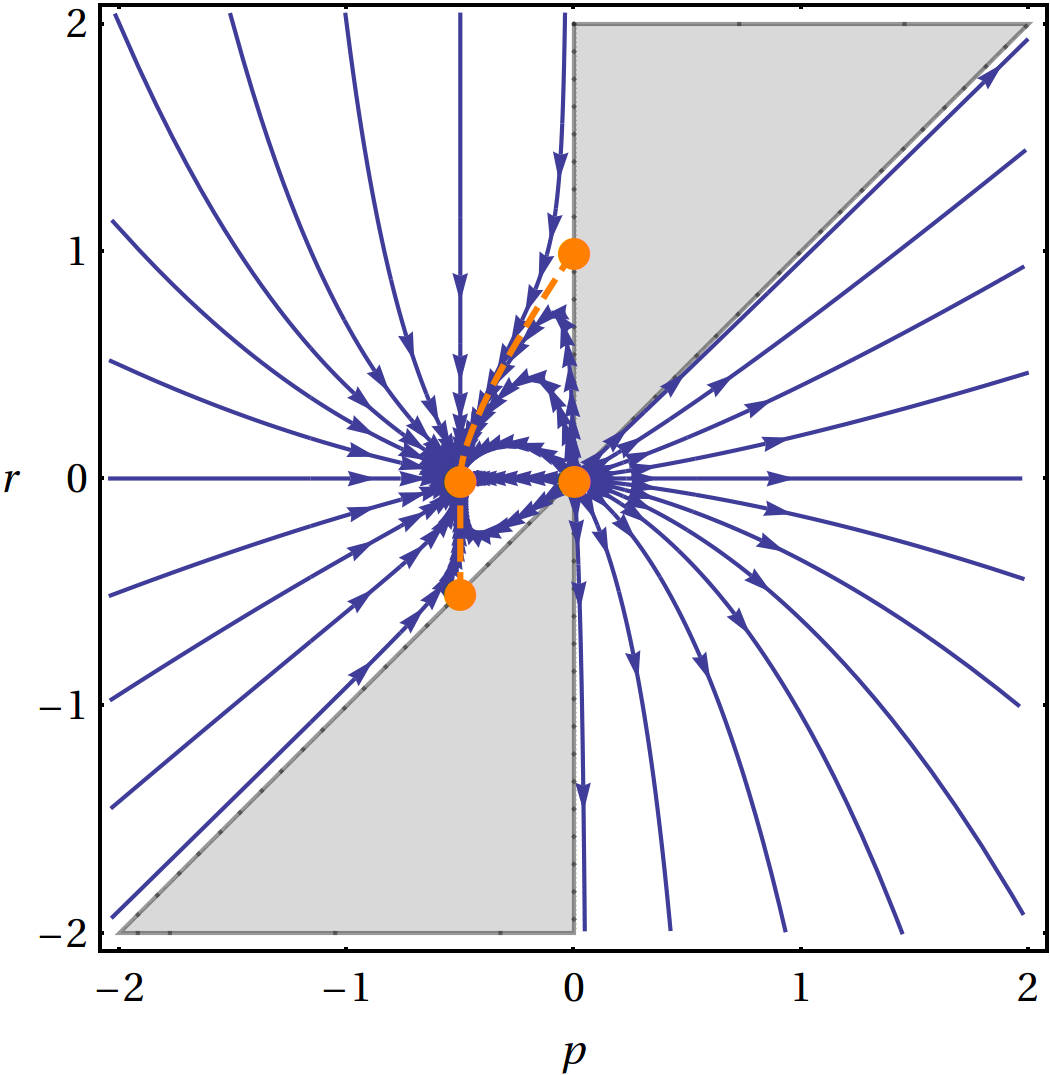}
\caption{Flow lines for $c < 0 < b$ (here $c = -1, b = 1$). The adiabatic ekpyrosis point forms a source at $(0,0)$, whereas the decelerated expansion point $(0,1)$ and the inflation point $(-\frac{1}{2}, -\frac{1}{2})$ are two saddle points connected by a separatrix. The tachyacoustic point $(-\frac{1}{2}, 0)$ on the separatrix forms a sink where the flow lines converge and $\epsilon$ approaches $\epsilon_* = 1$.} \label{fig:c=-1b=1}
\end{figure}

\begin{table}[h!]
\centering
\renewcommand{\arraystretch}{1.5}
\begin{tabular}{|c|c|c|c|c|c|c|}
\hline
$\alpha = b + c + 2$ & adia ekpy & \multicolumn{2}{|c|}{decel exp} & \multicolumn{2}{|c|}{inf / def} & tachyacoustic \\[-6pt]
$\beta = -c$ & $(0,0,2)$ & \multicolumn{2}{|c|}{$(0,1,2-c)$} & \multicolumn{2}{|c|}{$(\frac{-1}{b+1}, \frac{-1}{b+1}, \frac{b}{b+1})$} & $(\frac{c}{b-c}, \frac{b+c}{b-c}, \frac{2b}{b-c})$ \\
\hline
source & $\forall b, c$ & \multicolumn{2}{|c|}{$\diagup$} & \multicolumn{2}{|c|}{$\diagup$} & $\diagup$ \\
\hline
\multirow{2}{*}{saddle} & \multirow{2}{*}{$\diagup$} & \multicolumn{2}{|c|}{$c < 0$} & \multicolumn{2}{|c|}{$b < -1$ or $b > 0$} & \multirow{2}{*}{$b < 0 < c$} \\
\cline{3-6}
& & \multirow{2}{*}{$c = 0$} & $b < 0$ & \multirow{2}{*}{$b = 0$} & $c > 0$ & \\
\cline{1-2} \cline{4-4} \cline{6-7}
\multirow{2}{*}{sink} & \multirow{2}{*}{$\diagup$} & & $b \geq 0$ & & $c \leq 0$ & \multirow{2}{*}{$c < 0 < b$} \\
\cline{3-6}
& & \multicolumn{2}{|c|}{$c > 0$} & \multicolumn{2}{|c|}{$-1 < b < 0$} & \\
\hline
\end{tabular}
\caption{The fixed points in the model $\eta = c \, \epsilon + b$.} \label{tab:eta-epsilon}
\end{table}

As an example, consider the case $c < 0 < b$. The flow lines projected on the $p \, r$-plane is shown in figure~\ref{fig:c=-1b=1}.
There is a source point at $(0,0)$, two saddle points at $(0,1)$ and $(\frac{-1}{b+1}, \frac{-1}{b+1})$, connected by a separatrix that goes through a sink point at $(\frac{c}{b-c}, \frac{b+c}{b-c})$. In the right half plane, all trajectories emerge from the origin and flow away to infinity; the cosmic contraction near that point can be well described by the generalized adiabatic ekpyrotic model in section~\ref{sec:adek}. The left half plane is divided by the separatrix into two sectors. Trajectories in the right sector emerge from the origin and flow towards the sink, whereas trajectories to the left of the separatrix come from infinity and also flow to the same point. Hence in an expanding universe $\epsilon$ always approaches its value at the sink point, $\epsilon = (p-r)/p \big|_\text{sink} = -b/c = \epsilon_*$, as predicted by the running according to $\eta(\epsilon)$. The cosmic expansion near that point is described by the tachyacoustic model discussed in section~\ref{sec:tach}.

Similarly, an example for the case $-1 < b < 0 < c$ is shown in figure~\ref{fig:c=1,b=-1o3}.
\begin{figure}[t]
\centering
\includegraphics[width=3.5in]{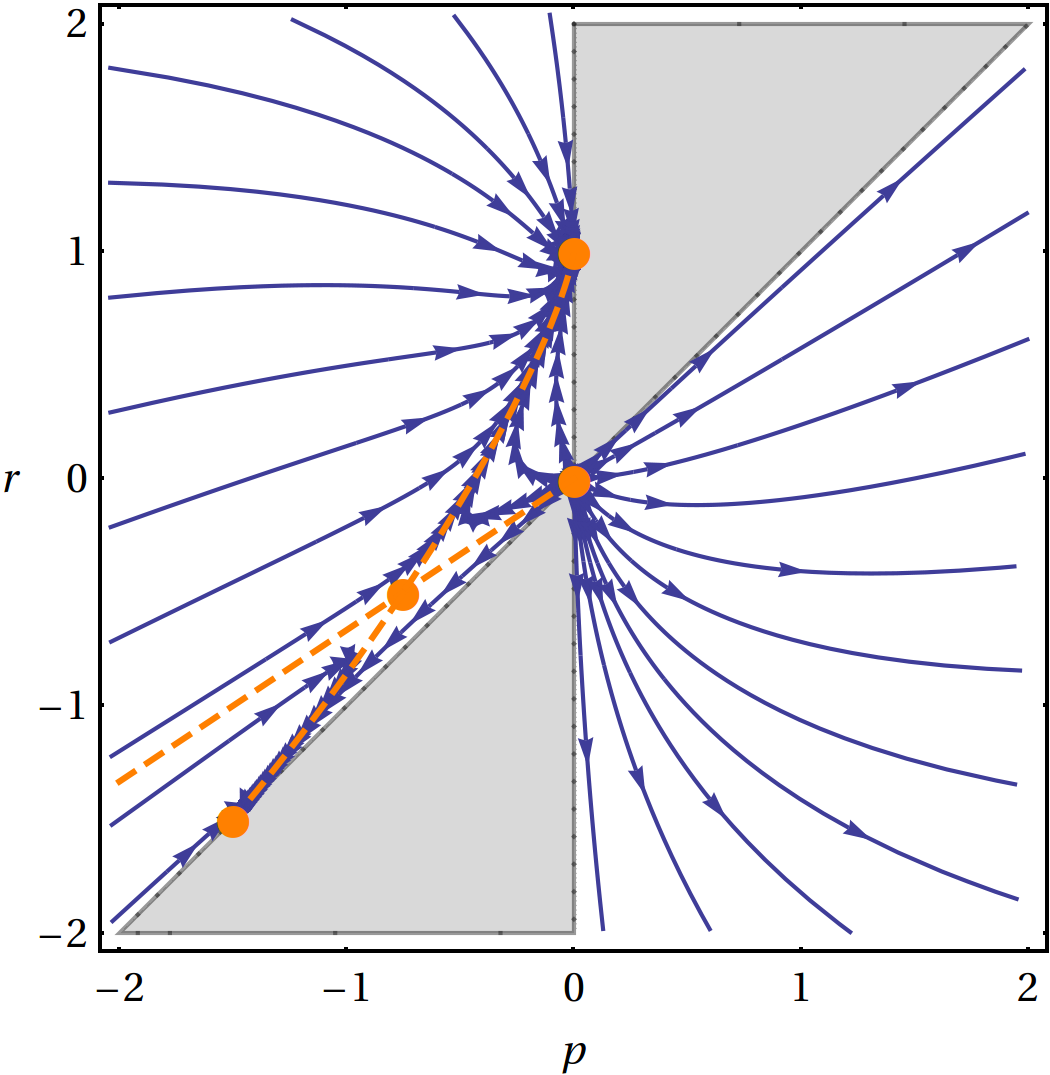}
\caption{Flow lines for $-1 < b < 0 < c$ (here $c = 1, b = -\frac{1}{3}$). Compared to figure~\ref{fig:c=-1b=1}, the adiabatic ekpyrosis point $(0,0)$ is still a source; both the decelerated expansion point $(0,1)$ and the inflation point $(-\frac{3}{2}, -\frac{3}{2})$ become sinks, whereas the tachyacoustic point $(-\frac{3}{4}, -\frac{1}{2})$ is no longer an attractor. The flow lines in the right half plane will approach the slope $1 - \epsilon_* = \frac{2}{3}$ further away from the origin.} \label{fig:c=1,b=-1o3}
\end{figure}
Here in the left half plane, the tachyacoustic point becomes a saddle point while the two endpoints of the separatrix become sinks instead, hence $\epsilon$ is repelled away from the value $\epsilon_*$ in an expanding universe. Meanwhile, in the right half plane, the trajectories emerge from the origin and eventually go parallel to the line with a slope $r/p = (b+c)/c = 1 - \epsilon_*$, proving that $\epsilon$ approaches $\epsilon_*$ in a contracting universe.

The case with $b < -1, 0 < c$ is even more different, as show in figure~\ref{fig:c=1,b=-2}.
\begin{figure}[t]
\centering
\includegraphics[width=3.5in]{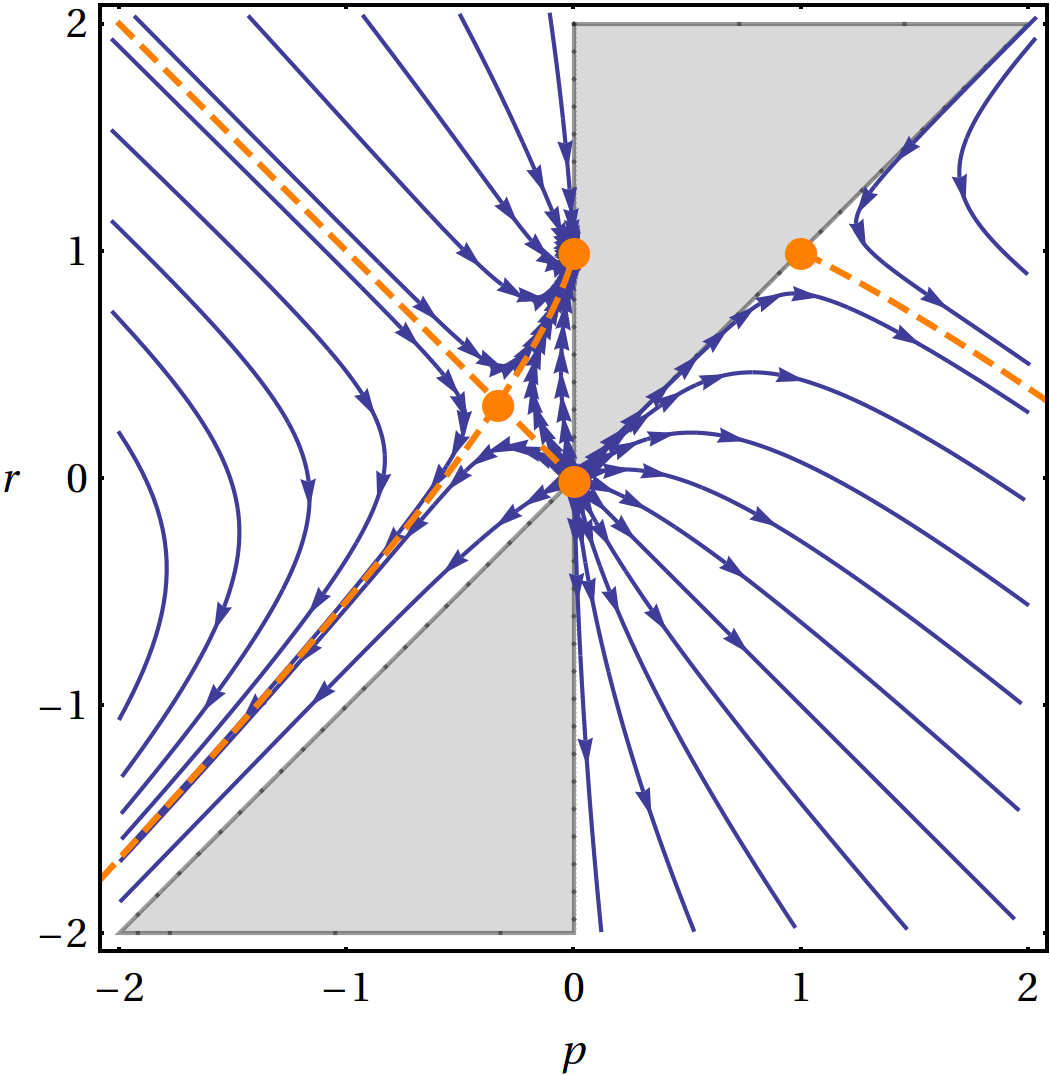}
\caption{Flow lines for $b < -1, 0 < c$ (here $c = 1, b = -2$). Compared to figure~\ref{fig:c=1,b=-1o3}, the inflation point in the left half plane is replaced by the deflation point in the right half plane, which is a saddle point at $(1,1)$. The flow lines in the right half plane approach the slope $1 - \epsilon_* = -1$ away from the origin.} \label{fig:c=1,b=-2}
\end{figure}
In this case the inflation point disappears from the left half plane; it is replaced by a saddle point on the border of $\epsilon = 0$ in the right half plane, attached with a separatrix that divides the right half plane into two sectors. This is the new situation discussed in section~\ref{sec:inf} --- near that saddle point, the cosmic contraction can be described by the deflation model with a decreasing speed of sound. Further away from that saddle point, the flow lines tend to approach the slope $r/p \to 1 - \epsilon_*$ just like in the case above; the cosmic contraction with an asymptotically constant $\epsilon = \epsilon_*$ can be described in the same way as the acoustic ekpyrotic model in section~\ref{sec:eps-cs}.

\section{Physical constraints} \label{sec:cons}

In the simplistic analysis above, each model can potentially produce a large number of exactly scale invariant modes. In practice, some physical constraints should be considered that could affect the stability of the models and limit the range of scale invariant modes \cite{Geshnizjani:2011dk, Baumann:2011dt}.

First of all, the period during which the scale invariant modes are created should be stable against the growth of spatial curvature and anisotropy. The effective densities of these two components scale as $1/a^2$ and $1/a^6$ respectively \cite{Erickson:2003zm}, whereas the dominant energy density of the universe scales as $1/a^{2\epsilon}$. Therefore an expansion phase with $\epsilon < 1$ can efficiently dilute away spatial curvature and anisotropy, leading to a flat, homogeneous and isotropic universe as observed today. Hence accelerated expansion models like inflation (with $\epsilon \ll 1$) and tachyacoustic expansion (with $\epsilon \lesssim 1$) can solve the horizon problem as well as provide a scale invariant power spectrum; but decelerated expansion models including adiabatic ekpyrotic expansion and apex (with $\epsilon \gg 1$) must be supplemented by additional mechanisms, e.g. an ekpyrotic contraction phase, to suppress inhomogeneity and anisotropy \cite{Joyce:2011ta}. Similarly, a slow contraction phase with $\epsilon > 3$, such as adiabatic ekpyrosis and acoustic ekpyrosis, can automatically beat curvature and anisotropy; but a rapid contraction such as deflation ($\epsilon \ll 1$) would require an extra stabilization phase, just like in the matter dominated contraction ($\epsilon = \frac{3}{2}$) model \cite{Cai:2012va}.

Besides, for models where the comoving Hubble horizon $1 / a H$ does not coincide with the freeze-out horizon, the scale invariant modes that have left the freeze-out horizon must also exit the Hubble horizon before the standard expansion phase. In models including adiabatic ekpyrosis, decelerated expansion, and deflation, this can be done by having a subsequent ekpyrotic phase to push the modes outside the Hubble horizon, such as in \cite{Khoury:2009my, Khoury:2011ii, Joyce:2011kh}.

Moreover, since the cosmological models assume classical general relativity, for consistency the energy density of the universe should remain sub-Planckian, $H^2 / M_\text{Pl}^2 \lesssim 1$. To estimate the size of the Hubble parameter when the scale invariant modes are generated, taking $n = -1$ in eq.~(\ref{eq:hankel}) to fix the coefficients in eq.~(\ref{eq:zeta_k}), one finds that the power spectrum (\ref{eq:power}) becomes
\begin{equation}
P_\zeta = \frac{1}{8 \pi^2 M_\text{Pl}^2 q^2 y^2} = \frac{1}{8 \pi^2 M_\text{Pl}^2} \frac{c_s}{\epsilon \, a^2 y^2} .
\end{equation}
From the measured amplitude of the power spectrum, $P_\zeta \approx 2.4 \times 10^{-9}$, the sub-Planckian energy density criterion can be expressed as
\begin{equation} \label{eq:H2/Mpl2}
\frac{H^2}{M_\text{Pl}^2} \approx \frac{8 \pi^2}{4 \times 10^8} \frac{\epsilon \, a^2 H^2 y^2}{c_s} \approx \frac{p^2 \epsilon \, c_s}{5 \times 10^6} \lesssim 1 .
\end{equation}

For models with time varying $\epsilon$ and $c_s$, it is also important to avoid strong coupling in generating scale invariant perturbations. A simple estimate and often reliable criterion is that the magnitude of the third order action for $\zeta$, $S_3 \sim \int \mathcal{O}(\zeta^3)$, should be smaller than the quadratic action $S_2$ in (\ref{eq:S2}) \cite{Leblond:2008gg, Baumann:2011dt}. For a rapidly growing $\epsilon$ or a rapidly decreasing $c_s$, the dominant contribution to the ratio $S_3 / S_2$ can be estimated by the term \cite{Joyce:2011kh}
\begin{equation} \label{eq:eps2/cs2}
\frac{S_3}{S_2} \sim \zeta \, \frac{\epsilon^2}{c_s^2} .
\end{equation}
This term results in a non-Gaussianity that peaks at small scales where the observational bound on $f_{NL} \sim S_3 / \zeta \, S_2$ is weak. Demanding that $S_3 / S_2 \lesssim 1$ when the scale invariant modes crosses the horizon, the weak coupling criterion can be written as
\begin{equation}
\frac{\epsilon^2}{c_s^2} \lesssim \frac{1}{\zeta} \sim \frac{1}{5\times10^{-5}} .
\end{equation}

The above two constraints can quantitatively limit the number of scale invariant modes generated in certain types of models. For example, in the tachyacoustic model with a constant $\epsilon \sim \mathcal{O}(1)$, using $p = \tilde{s}/2 - 1 = -1/(\epsilon+1)$ from section~\ref{sec:tach}, eq.~(\ref{eq:H2/Mpl2}) becomes $c_s \epsilon / (\epsilon+1)^2 \lesssim 5 \times 10^6$. Combining that with eq.~(\ref{eq:eps2/cs2}), one obtains
\begin{equation} \label{eq:tachycs}
7\times10^{-3} \, \epsilon \lesssim c_s \lesssim 5\times10^6 \, \frac{(\epsilon+1)^2}{\epsilon} .
\end{equation}
Since $c_s \sim (-y)^{\tilde{s}} \sim k^{-2\epsilon/(\epsilon+1)}$ at horizon crossing, the number of scale invariant modes is bounded by \cite{Joyce:2011kh}
\begin{equation}
N \lesssim \frac{\epsilon+1}{2\epsilon} \log \bigg( \frac{5\times10^6}{7\times10^{-3}} \Big( \frac{\epsilon+1}{\epsilon} \Big)^2 \bigg) \approx 22 .
\end{equation}
Similarly, in the decelerated expansion model described in section~\ref{sec:apex}, using $\epsilon \approx -1/p$, eqs.~(\ref{eq:H2/Mpl2}) and (\ref{eq:eps2/cs2}) imply $7\times10^{-3} \lesssim c_s / \epsilon \lesssim 5\times10^6$. Since in this model $c_s / \epsilon \sim y^2 \sim 1/k^2$ at horizon crossing, the number of scale invariant modes is bound by $N \lesssim \frac{1}{2} \log (5\times10^6 / 7\times10^{-3}) \approx 10$, just enough to encompass the observed 10 e-folds of primordial density fluctuations. This result is the same as for the apex model \cite{Khoury:2010gw}.

Finally, models with a time varying $c_s$ often run into a superluminal speed of sound, $c_s > 1$. Such a superluminal speed of sound does not necessarily violate the causal structure of spacetime \cite{Bruneton:2006gf, Babichev:2007dw}. Nevertheless, it is important to understand how constrained the cosmological models become if $c_s \leq 1$. It is can be seen from eq.~(\ref{eq:tachycs}) that the tachyacoustic model requires superluminality (hence the name) for generating a wide range of scale invariant modes; otherwise this constraint becomes $7\times10^{-3} \, \epsilon \lesssim c_s \leq 1$, allowing only $N \lesssim \frac{1}{2} \log(10^3 / 7 \epsilon) \approx 3$ e-folds of scale invariant modes for $\epsilon \sim \mathcal{O}(1)$. The same bound also applies to the particular adiabatic ekpyrotic model described in section~\ref{sec:adek} and the acoustic ekpyrotic model described in section~\ref{sec:eps-cs}. Similarly, for the decelerated expansion model discussed above, since $\epsilon \gg 1$, one finds $7\times10^{-3} \lesssim c_s / \epsilon \ll 1$, posing an even tighter bound on $N$. Hence these models all rely on having a superluminal speed of sound for generating a sufficient number of scale invariant modes, in the same spirit as the tachyacoustic model.

\section{Summary} \label{sec:summ}

We have presented a set of flow parameters that are used to depict single field cosmological models. Scale invariant perturbations are produced in the model if these flow parameters satisfy specific dynamical equations. We have analyzed the flow of these parameters by identifying the fixed points and their properties. Besides existing models including inflation, adiabatic ekpyrosis, apex, and tachyacoustic expansion, four new scenarios have emerged from our analysis:
\begin{itemize}
\item A generalization of the adiabatic ekpyrotic model with a time varying speed of sound, described in section~\ref{sec:adek}. In a particular model, the scale invariance of the power spectrum completely relies on the time dependence of $c_s$.
\item An extremely decelerated expansion with a decreasing speed of sound, described in section~\ref{sec:apex}.
\item An exponentially rapid contraction, or ``deflation'', described in section~\ref{sec:inf}. Scale invariant modes are generated when $\epsilon$ increases and $c_s$ decreases exponentially.
\item An ``acoustic ekpyrotic'' contraction, described in section~\ref{sec:eps-cs}. Based on the usual ekpyrotic contraction, it requires a particularly time dependent speed of sound.
\end{itemize}

These new models serve as distinctive examples in which both $\epsilon$ and $c_s$ vary with time cooperatively to create scale invariant perturbations. The range of scale invariant modes can be estimated by considering physical constraints from sub-Planckian energy density and weak coupling of higher order perturbations. Unless subluminality is imposed, all these models can adequately account for the observed range of the scale invariant power spectrum.

The flow parameters that we used to find the new cosmological models are convenient for obtaining the scale invariant power spectrum as well as analyzing the physical constraints. We note that in \cite{Geshnizjani:2011rm} a different set of flow parameters are defined by expanding the Hubble parameter $H$ as a function of the field $\phi$. Those flow parameters can be used to effectively reconstruct scalar field models, but these cosmological models do not acquire scale invariant perturbations. It may be interesting to relate the two sets of flow parameters in order to construct single field models that implement the new mechanisms for generating a scale invariant power spectrum.

\bigskip
\noindent
\textbf{Acknowledgments}

\bigskip
\noindent
I thank Adam Brown, Enrico Pajer, and Paul Steinhardt for enormously helpful discussions.

\appendix
\addtocontents{toc}{\protect\setcounter{tocdepth}{1}}

\section{$\eta$ as a function of $(\epsilon, s)$} \label{sec:A}

Consider a scalar field $\phi$ with the Lagrangian $\mathcal{L} = P(\phi, X)$, where $X \equiv -\frac{1}{2}(\partial \phi)^2$. The homogeneous evolution on a flat Friedmann-Robertson-Walker background is determined by the equations of motion
\begin{align} \label{eq:eom}
\dot{\phi} &= \sqrt{2X} , \\
\dot{X} &= - 6 c_s^2 H X - \rho_{,\phi} \sqrt{2X} / \rho_{,X} .
\end{align}
Here the energy density $\rho(\phi, X) \equiv 2 X P_{,X} - P$, and
\begin{equation}
H = \pm \sqrt{\rho / 3} \, , \quad c_s^2 = P_{,X} / \rho_{,X} .
\end{equation}
where the $\pm$ sign corresponds to the universe being expanding/contracting.

The flow parameters $\epsilon$, $s$ and $\eta$ are defined by
\begin{equation}
\epsilon \equiv \frac{-\dot{H}}{H^2} , \quad s \equiv \frac{\dot{c_s}}{H c_s} , \quad \eta \equiv \frac{\dot{\epsilon}}{H \epsilon} .
\end{equation}
Using the equations of motion, they can be explicitly written as functions of $(\phi, X)$,
\begin{align}
\epsilon &= \frac{3 X P_{,X}}{2 X P_{,X} - P} \ , \label{eq:eps(phi,X)} \\[4pt]
s &= \frac{6X (P_{,X} P_{,XX} - X P_{,XX}^2 + X P_{,X} P_{,XXX})}{(2X P_{,XX} + P_{,X})^2} \pm \sqrt{\frac{6X}{2X P_{,X} - P}} \; \times \label{eq:s(phi,X)} \\
&\quad \left[ \frac{(P_{,X} P_{,XX} - X P_{,XX}^{\;2} + X P_{,X} P_{,XXX}) (2X P_{,X\phi} - P_{,\phi})}{P_{,X} (2X P_{,XX} + P_{,X})^2} + \frac{X (P_{,XX} P_{,X\phi} - P_{,X} P_{,XX\phi})}{P_{,X} (2X P_{,XX} + P_{,X})} \right] , \nonumber \\[4pt]
\eta &= \frac{6X P_{,X}}{2X P_{,X} - P} - \frac{6 (P_{,X} + X P_{,XX})}{2X P_{,XX} + P_{,X}} \pm \frac{(P_{,X} P_{,\phi} + X P_{,XX} P_{,\phi} - X P_{,X} P_{,X\phi})}{X P_{,X} (2X P_{,XX} + P_{,X})} \sqrt{\frac{6X}{2X P_{,X} - P}} \ . \label{eq:eta(phi,X)}
\end{align}

Here we argue that each model $P(\phi, X)$ determines a particular function $\eta(\epsilon, s)$. Indeed, for a given function $P(\phi, X)$, eqs.~(\ref{eq:eps(phi,X)}, \ref{eq:s(phi,X)}) can be locally inverted to express $(\phi, X)$ in terms of $(\epsilon, s)$, which can then be used in eq.~(\ref{eq:eta(phi,X)}) to express $\eta$ as a function of $(\epsilon, s)$, at least locally. Below we consider a few examples.

\subsection{$P(\phi, X) = K(X) - V(\phi)$}

For scalar models with $P(\phi, X) = K(X) - V(\phi)$, eqs.~(\ref{eq:eps(phi,X)}, \ref{eq:s(phi,X)}, \ref{eq:eta(phi,X)}) become
\begin{align}
\epsilon &= \frac{3 X K_{,X}}{2 X K_{,X} - K + V} \ , \\[4pt]
s &= \frac{6X (K_{,X} K_{,XX} - X K_{,XX}^2 + X K_{,X} K_{,XXX})}{(2X K_{,XX} + K_{,X})^2} \left[ 1 \pm \frac{V_{,\phi}}{6X K_{,X}} \sqrt{\frac{6X}{2X K_{,X} - K + V}} \; \right] , \\[4pt]
\eta &= \frac{6X K_{,X}}{2X K_{,X} - K + V} - \frac{6 (K_{,X} + X K_{,XX})}{2X K_{,XX} + K_{,X}} \left[ 1 \pm \frac{V_{,\phi}}{6X K_{,X}} \sqrt{\frac{6X}{2X K_{,X} - K + V}} \; \right] .
\end{align}

For a canonical scalar field we have $K(X) = X$. Then those parameters are given by
\begin{equation}
\epsilon = \frac{3X}{X + V} \ , \quad s = 0 , \quad \eta = \frac{6X}{X + V} - 6 \mp \frac{V_{,\phi}}{X} \sqrt{\frac{6X}{X + V}} .
\end{equation}
This case is generically degenerate since $s$ is trivial and $(\phi, X)$ cannot be solved from $\epsilon$ alone. However, there is a special case where $V$ is exponential, $V \propto e^{c \, \phi}$. In that case we have $V_{,\phi} / X = c V / X = c (3 - \epsilon) / \epsilon$, hence $\eta$ can be expressed as a function of $\epsilon$,
\begin{equation}
\eta(\epsilon) = 2 (\epsilon - 3) \bigg( 1 \pm \frac{c}{\sqrt{2\epsilon}} \bigg) .
\end{equation}
For a constant potential $V = V_0$, for example, we find a linear relation $\eta(\epsilon) = 2 \epsilon - 6$.

\subsection{$P(\phi, X) = F(\phi) \bar{P}(X)$}

For another type of models with $P(\phi, X) = F(\phi) \bar{P}(X)$, we have
\begin{align}
\epsilon &= \frac{3 X \bar{P}_{,X}}{2 X \bar{P}_{,X} - \bar{P}} \ , \\[4pt]
s &= \frac{6X (\bar{P}_{,X} \bar{P}_{,XX} - X \bar{P}_{,XX}^2 + X \bar{P}_{,X} \bar{P}_{,XXX})}{(2X \bar{P}_{,XX} + \bar{P}_{,X})^2} \left[ 1 \pm \frac{F_{,\phi}}{F^{3/2}} \sqrt{\frac{2X \bar{P}_{,X} - \bar{P}}{6X \bar{P}_{,X}^2}} \; \right] , \\[4pt]
\eta &= \frac{6 (\bar{P} \bar{P}_{,X} - X \bar{P}_{,X}^2 + X \bar{P} \bar{P}_{,XX})}{(2X \bar{P}_{,X} - \bar{P}) (2X \bar{P}_{,XX} + \bar{P}_{,X})} \left[ 1 \pm \frac{F_{,\phi}}{F^{3/2}} \sqrt{\frac{2X \bar{P}_{,X} - \bar{P}}{6X \bar{P}_{,X}^2}} \; \right] .
\end{align}
In this case $\epsilon$ is a function of $X$ only, and {\it vice versa}, at least locally. Since the factor in the brackets are the same for $s$ and $\eta$, and the other factors are functions of $X$ only, we can write
\begin{equation}
\eta = f(\epsilon) \, s ,
\end{equation}
where the function $f(\epsilon)$ is determined by the function $\bar{P}(X)$. For example, with $\bar{\rho}(X) \equiv 2X \bar{P}_{,X} - \bar{P} \propto X^\gamma$ $(\gamma \neq \frac{1}{2})$, we find $f(\epsilon) = 2$, and hence $\eta = 2 s$.

A special case is where $F(\phi) \propto \phi^{-2}$, then we have $F_{,\phi} / F^{3/2} =$~const, hence both $s$ and $\eta$ depend only on $X$. Therefore $\eta$ can be expressed as a function of either $\epsilon$ or $s$ alone.

\end{document}